\documentclass[12pt]{article}
\usepackage{amsmath,amsthm}
\usepackage{eufrak}
\usepackage{pb-diagram,lamsarrow,pb-lams}

\newcommand{\ar}{\renewcommand{\arraystretch}{1}} % 1.0 % 0.6
\DeclareMathAlphabet{\bb}{U}{msb}{m}{n} \gdef\C{\bb C} \gdef\dZ{\bb
Z}   \gdef\dS{\bb S}
\gdef\R{\bb R} \gdef\K{\bb K} \gdef\BH{\bb H} \gdef\F{\bb F}

\DeclareMathOperator{\End}{End} \DeclareMathOperator{\spin}{{\bf
Spin}} \DeclareMathOperator{\pin}{{\bf Pin}}
\DeclareMathOperator{\fD}{\mathfrak{D}}
\DeclareMathOperator{\Id}{Id} 
\DeclareMathOperator{\Aut}{Aut} \DeclareMathOperator{\sAut}{{\sf
Aut}} \DeclareMathOperator{\sExt}{{\sf Ext}}

\DeclareMathOperator{\Sym}{Sym} 
 
\DeclareMathOperator{\Ext}{Ext} 
 
 \DeclareMathOperator{\SO}{SO}
\DeclareMathOperator{\SL}{SL} \DeclareMathOperator{\SU}{SU}
\DeclareMathOperator{\Sp}{Sp}\DeclareMathOperator{\GO}{O}

\newcommand{\cA}{\mathcal{A}}

\newcommand{\cE}{\mathcal{E}}

\newcommand{\sA}{{\sf A}}
\newcommand{\sB}{{\sf B}}
\newcommand{\sI}{{\sf I}}
\newcommand{\sW}{{\sf W}}
\newcommand{\sE}{{\sf E}}
\newcommand{\sC}{{\sf C}}
\newcommand{\sF}{{\sf F}}
\newcommand{\sT}{{\sf T}}
\newcommand{\sS}{{\sf S}}

\newcommand{\sX}{{\sf X}}
\newcommand{\sY}{{\sf Y}}
\newcommand{\sK}{{\sf K}}

\newcommand{\bi}{{\bf i}}

\newcommand{\bx}{{\bf x}}

\newcommand{\bz}{{\bf z}}
\newcommand{\balpha}{\boldsymbol{\alpha}}

\newcommand{\fG}{\mathfrak{G}}
\newcommand{\fC}{\mathfrak{C}}

\newcommand{\fc}{\mathfrak{c}}
\newcommand{\fg}{\mathfrak{g}}

\newcommand{\fq}{\mathfrak{q}}

\newcommand{\Lip}{\boldsymbol{\Gamma}}

\newcommand{\cl}{C\kern -0.2em \ell}

\newcommand{\p}{\prime}
\newcommand{\e}{\mbox{\bf e}}

\newcommand{\ld}{\left[}
\newcommand{\rd}{\right]}

\newtheorem{theorem}{Theorem}

\newtheorem{prop}{Proposition}

\begin{document}
\title{$CPT$ groups of higher spin fields}
\author{V.~V. Varlamov\thanks{Siberian State Industrial University,
Kirova 42, Novokuznetsk 654007, Russia, e-mail:
vadim.varlamov@mail.ru}}
\date{}
\maketitle
\begin{abstract}
$CPT$ groups of higher spin fields are defined in the framework of
automorphism groups of Clifford algebras associated with the complex
representations of the proper orthochronous Lorentz group. Higher
spin fields are understood as the fields on the Poincar\'{e} group
which describe orientable (extended) objects. A general method for
construction of $CPT$ groups of the fields of any spin is given.
$CPT$ groups of the fields of spin-1/2, spin-1 and spin-3/2 are
considered in detail. $CPT$ groups of the fields of tensor type are
discussed. It is shown that tensor fields correspond to particles of
the same spin with different masses.
\end{abstract}
%MSC 2000:\;{\bf 15A66, 15A90, 20645}\\

\section{Introduction}
In 2003, $CPT$ group was introduced \cite{Var04a} in the context of
an extension of automorphism groups of Clifford algebras. The
relationship between $CPT$ groups and extraspecial groups and
universal coverings of orthogonal groups was established in
\cite{Var04a,Var04b}. In 2004, Socolovsky considered the $CPT$ group
of the spinor field with respect to phase quantities \cite{Soc04}
(see also \cite{CCPS06,CS09a,CS09b,CS10,Pla09}). $CPT$ groups of
spinor fields in the de Sitter spaces of different signatures were
studied in the works \cite{Var05c,Var05}. The following logical step
in this direction is a definition of the $CPT$ groups for the higher
spin fields. The formalism developed in the previous works
\cite{Var04a,Var04b} allows us to define $CPT$ groups for the fields
of any spin on the spinspaces associated with representations of the
spinor group $\spin_+(1,3)$ (a universal covering of the proper
orthochronous Lorentz group).

Our consideration based on the concept of generalized wavefunctions
introduced by Ginzburg and Tamm in 1947 \cite{GT47}, where the
wavefunction depends both coordinates $x_\mu$ and additional
internal variables $\theta_\mu$ which describe spin of the particle,
$\mu=0,1,2,3$. In 1955, Finkelstein showed \cite{Fin55} that
elementary particles models with internal degrees of freedom can be
described on manifolds larger then Minkowski spacetime (homogeneous
spaces of the Poincar\'{e} group). The quantum field theories on the
Poincar\'{e} group were discussed in the papers
\cite{Lur64,Kih70,BF74,Aro76,KLS95,Tol96,LSS96,Dre97,GS01,GL01}. A
consideration of the field models on the homogeneous spaces leads
naturally to a generalization of the concept of wave function
(fields on the Poincar\'{e} group). The general form of these fields
is related closely with the structure of the Lorentz and
Poincar\'{e} group representations \cite{GMS,Nai58,BBTD88,GS01} and
admits the following factorization $f(x,\bz)=\phi^n(\bz)\psi_n(x)$,
where $x\in T_4$ and $\phi^n(\bz)$ form a basis in the
representation space of the Lorentz group. At this point, four
parameters $x^\mu$ correspond to position of the point-like object,
whereas remaining six parameters $\bz\in\spin_+(1,3)$ define
orientation in quantum description of orientable (extended) object
\cite{GS09,GS10} (see also \cite{Kai09}). It is obvious that the
point-like object has no orientation, therefore, orientation is an
intrinsic property of the extended object. On the other hand,
measurements in quantum field theory lead to extended objects. As is
known, loop divergences emerging in the Green functions in quantum
field theory originate from correspondence of the Green functions to
\emph{unmeasurable} (and hence unphysical) point-like quantities.
This is because no physical quantity can be measured in a point, but
in a region, the size of which (or 'diameter' of the extended
object) is constrained by the resolution of measuring equipment
\cite{Alt10}. Taking it into account, we come to consideration of
physical quantity as an extended object, the generalized
wavefunction of which is described by the field
\[
\boldsymbol{\psi}(\balpha)=\langle x,g\,|\boldsymbol{\psi}\rangle
\]
on the homogeneous space of some orthogonal group $\SO(p,q)$, where
$x\in T_n$ (position) and $g\in\spin_+(p,q)$ (orientation), $n=p+q$.
So, in \cite{SZ92,SZ94} Segal and Zhou proved convergence of quantum
field theory, in particular, quantum electrodynamics, on the
homogeneous space $R^1\times S^3$ of the conformal group $\SO(2,4)$,
where $S^3$ is the three-dimensional real sphere.

In the present work we describe discrete symmetries of the
generalized wavefunctions $\boldsymbol{\psi}(\balpha)=\langle
x,\fg\,|\boldsymbol{\psi}\rangle$ (fields on the Poincar\'{e} group)
in terms of involutive automorphisms of the subgroup $\spin_+(p,q)$,
As is known, the universal covering of the proper Poincar\'{e} group
is isomorphic to a semidirect product $\SL(2;\C)\odot T_4$ or
$\spin_+(1,3)\odot T_4$. Since the group $T_4$ is Abelian, then all
its representations are one-dimensional. Thus, all the
finite-dimensional representations of the proper Poincar\'{e} group
in essence are equivalent to the representations $\fC$ of the group
$\spin_+(1,3)$.

An algebraic method for description of discrete symmetries was
proposed by author in the works \cite{Var99,Var00,Var04a,Var04b},
where the discrete symmetries are represented by fundamental
automorphisms of the Clifford algebras. So, the space inversion $P$,
time reversal $T$ and their combination $PT$ correspond to an
automorphism $\star$ (involution), an antiautomorphism
$\widetilde{\phantom{cc}}$ (reversion) and an antiautomorphism
$\widetilde{\star}$ (conjugation), respectively. The fundamental
automorphisms of the Clifford algebras are compared to elements of
the finite group formed by the discrete transformations. In turn, a
set of the fundamental automorphisms, added by an identical
automorphism, forms a finite group $\Aut(\cl)$, for which in virtue
of the Wedderburn-Artin Theorem there exists a matrix (spinor)
representation. Further, other important discrete symmetry is the
charge conjugation $C$. In contrast with the transformations $P$,
$T$, $PT$, the operation $C$ is not space-time discrete symmetry.
This transformation is firstly appeared on the representation spaces
of the Lorentz group and its nature is strongly different from other
discrete symmetries. For that reason the charge conjugation $C$ is
represented by a pseudoautomorphism $\cA\rightarrow\overline{\cA}$
which is not fundamental automorphism of the Clifford algebra. All
spinor representations of the pseudoautomorphism
$\cA\rightarrow\overline{\cA}$ were given in \cite{Var04a}. An
introduction of the transformation $\cA\rightarrow\overline{\cA}$
allows us to extend the automorphism group $\Aut(\cl)$ of the
Clifford algebra. It was shown \cite{Var04a} that automorphisms
$\cA\rightarrow\cA^\star$, $\cA\rightarrow\widetilde{\cA}$,
$\cA\rightarrow\widetilde{\cA^\star}$,
$\cA\rightarrow\overline{\cA}$,
$\cA\rightarrow\overline{\cA^\star}$,
$\cA\rightarrow\overline{\widetilde{\cA}}$ and
$\cA\rightarrow\overline{\widetilde{\cA^\star}}$ form a finite group
of order 8 (an extended automorphism group $\Ext(\cl)=\{\Id,\star,
\widetilde{\phantom{cc}},\widetilde{\star},\overline{\phantom{cc}},
\overline{\star},\overline{\widetilde{\phantom{cc}}},
\overline{\widetilde{\star}}\}$). The group $\Ext(\cl)$ is a
generating group of the full $CPT$ group $\{\pm 1,\pm P,\pm T,\pm
PT,\pm C,\pm CP,\pm CT,\pm CPT\}$. There are also other realizations
of the discrete symmetries via the automorphisms of the Lorentz and
Poincar\'{e} groups, see \cite{GMS,Mic64,Kuo71,Sil92,BGS00}.

The present paper is organized as follows. In the section 2 we
briefly discuss the basis notions concerning Clifford algebras and
$CPT$ groups, and also we consider their descriptions within
universal coverings of orthogonal groups and spinor representations.
In the section 3 we introduce the main objects of our study, $CPT$
groups of higher spin fields. These groups are defined on the system
of complex representations of the group $\spin_+(1,3)$. In the
sections 4--6 we consider in detail $CPT$ groups for the fields
$(1/2,0)\oplus(0,1/2)$, $(1,0)\oplus(0,1)$ and
$(3/2,0)\oplus(0,3/2)$. In the section 7 we define $CPT$ groups for
the fields of tensor type.

\section{Algebraic and group theoretical preliminaries}
In this section we will consider some basic facts concerning
automorphisms of the Clifford algebras and universal coverings of
orthogonal groups.

Let $\F$ be a field of characteristic 0 $(\F=\R,\,\F=\C)$,
\index{field!of characteric 0} where $\R$ and $\C$ are the fields of
real and complex numbers, respectively. A Clifford algebra $\cl$
over a field $\F$ is an algebra with $2^n$ basis
elements\index{element!basis}: $\e_0$ (unit of the algebra)
$\e_1,\e_2,\ldots,\e_n$ and products of the one--index elements
$\e_{i_1i_2\ldots i_k}=\e_{i_1}\e_{i_2}\ldots\e_{i_k}$. Over the
field $\F=\R$ the Clifford algebra is denoted as $\cl_{p,q}$, where
the indices $p$ and $q$ correspond to the indices of the quadratic
form\index{form!quadratic}
\[
Q=x^2_1+\ldots+x^2_p-\ldots-x^2_{p+q}
\]
of a vector space $V$ associated with $\cl_{p,q}$.

An arbitrary element $\cA$ of the algebra $\cl_{p,q}$ is represented
by a following formal polynomial:
\begin{multline}
\cA=a^0\e_0+\sum^n_{i=1}a^i\e_i+\sum^n_{i=1}\sum^n_{j=1}a^{ij}\e_{ij}+
\ldots+\sum^n_{i_1=1}\cdots\sum^n_{i_k=1}a^{i_1\ldots
i_k}\e_{i_1\ldots i_k}+\\
+\ldots+a^{12\ldots n}\e_{12\ldots n}=\sum^n_{k=0}a^{i_1i_2\ldots
i_k} \e_{i_1i_2\ldots i_k}.\nonumber
\end{multline}

In Clifford algebra $\cl$ there exist four fundamental automorphisms.\\[0.2cm]
1) {\bf Identity}: An automorphism $\cA\rightarrow\cA$ and
$\e_{i}\rightarrow\e_{i}$.\\
This automorphism is an identical automorphism of the algebra $\cl$.
$\cA$ is an arbitrary element of $\cl$.\\[0.2cm]
2) {\bf Involution}: An automorphism $\cA\rightarrow\cA^\star$ and
$\e_{i}\rightarrow-\e_{i}$.\\
In more details, for an arbitrary element $\cA\in\cl$ there exists a
decomposition $ \cA=\cA^{\p}+\cA^{\p\p}, $ where $\cA^{\p}$ is an
element consisting of homogeneous odd elements, and $\cA^{\p\p}$ is
an element consisting of homogeneous even elements, respectively.
Then the automorphism $\cA\rightarrow\cA^{\star}$ is such that the
element $\cA^{\p\p}$ is not changed, and the element $\cA^{\p}$
changes sign: $ \cA^{\star}=-\cA^{\p}+\cA^{\p\p}. $ If $\cA$ is a
homogeneous element, then
\begin{equation}\label{auto16}
\cA^{\star}=(-1)^{k}\cA,
\end{equation}
where $k$ is a degree of the element. It is easy to see that the
automorphism $\cA\rightarrow\cA^{\star}$ may be expressed via the
volume element $\omega=\e_{12\ldots p+q}$:
\begin{equation}\label{auto17}
\cA^{\star}=\omega\cA\omega^{-1},
\end{equation}
where $\omega^{-1}=(-1)^{\frac{(p+q)(p+q-1)}{2}}\omega$. When $k$ is
odd, the basis elements $\e_{i_{1}i_{2}\ldots i_{k}}$ the sign
changes, and when $k$ is even, the sign
is not changed.\\[0.2cm]
3) {\bf Reversion}: An antiautomorphism
$\cA\rightarrow\widetilde{\cA}$ and
$\e_i\rightarrow\e_i$.\\
The antiautomorphism $\cA\rightarrow\widetilde{\cA}$ is a reversion
of the element $\cA$, that is the substitution of each basis element
$\e_{i_{1}i_{2}\ldots i_{k}}\in\cA$ by the element
$\e_{i_{k}i_{k-1}\ldots i_{1}}$:
\[
\e_{i_{k}i_{k-1}\ldots i_{1}}=(-1)^{\frac{k(k-1)}{2}}
\e_{i_{1}i_{2}\ldots i_{k}}.
\]
Therefore, for any $\cA\in\cl_{p,q}$ we have
\begin{equation}\label{auto19}
\widetilde{\cA}=(-1)^{\frac{k(k-1)}{2}}\cA.
\end{equation}
4) {\bf Conjugation}: An antiautomorphism
$\cA\rightarrow\widetilde{\cA^\star}$
and $\e_i\rightarrow-\e_i$.\\
This antiautomorphism is a composition of the antiautomorphism
$\cA\rightarrow\widetilde{\cA}$ with the automorphism
$\cA\rightarrow\cA^{\star}$. In the case of a homogeneous element
from the formulae (\ref{auto16}) and (\ref{auto19}), it follows
\begin{equation}\label{20}
\widetilde{\cA^{\star}}=(-1)^{\frac{k(k+1)}{2}}\cA.
\end{equation}
As is known, the complex algebra $\C_n$ is associated with a complex
vector space $\C^n$. Let $n=p+q$, then an extraction operation of
the real subspace $\R^{p,q}$ in $\C^n$  forms the foundation of
definition of the discrete transformation known in physics as {\it a
charge conjugation} $C$. Indeed, let $\{\e_1,\ldots,\e_n\}$ be an
orthobasis in the space $\C^n$, $\e^2_i=1$. Let us remain the first
$p$ vectors of this basis unchanged, and other $q$ vectors multiply
by the factor $i$. Then the basis
\begin{equation}\label{6.23}
\left\{\e_1,\ldots,\e_p,i\e_{p+1},\ldots,i\e_{p+q}\right\}
\end{equation}
allows one to extract the subspace $\R^{p,q}$ in $\C^n$. Namely, for
the vectors $\R^{p,q}$ we take the vectors of $\C^n$ which decompose
on the basis (\ref{6.23}) with real coefficients. In such a way we
obtain a real vector space $\R^{p,q}$ endowed (in general case) with
a non--degenerate quadratic form\index{form!quadratic}
\[
Q(x)=x^2_1+x^2_2+\ldots+x^2_p-x^2_{p+1}-x^2_{p+2}-\ldots-x^2_{p+q},
\]
where $x_1,\ldots,x_{p+q}$ are coordinates of the vector $\bx$ in
the basis (\ref{6.23}). It is easy to see that the extraction of
$\R^{p,q}$ in $\C^n$ induces an extraction of {\it a real
subalgebra}\index{subalgebra!real} $\cl_{p,q}$ in $\C_n$. Therefore,
any element $\cA\in\C_n$ can be unambiguously represented in the
form
\[
\cA=\cA_1+i\cA_2,
\]
where $\cA_1,\,\cA_2\in\cl_{p,q}$. The one-to-one mapping
\begin{equation}\label{6.24}
\cA\longrightarrow\overline{\cA}=\cA_1-i\cA_2
\end{equation}
transforms the algebra $\C_n$ into itself with preservation of
addition and multiplication operations for the elements $\cA$; the
operation of multiplication of the element $\cA$ by the number
transforms to an operation of multiplication by the complex
conjugate number. Any mapping of $\C_n$ satisfying these conditions
is called {\it a pseudoautomorphism}.\index{pseudoautomorphism}
Thus, the extraction of the subspace $\R^{p,q}$ in the space $\C^n$
induces in the algebra $\C_n$ a pseudoautomorphism
$\cA\rightarrow\overline{\cA}$ \cite{Ras55,Ras58}.

An introduction of the pseudoautomorphism
$\cA\rightarrow\overline{\cA}$ allows us to extend the automorphism
set of the complex Clifford algebra $\C_n$. Namely, we add to the
four fundamental automorphisms $\cA\rightarrow\cA$,
$\cA\rightarrow\cA^\star$, $\cA\rightarrow\widetilde{\cA}$,
$\cA\rightarrow\widetilde{\cA^\star}$ the pseudoautomorphism
$\cA\rightarrow\overline{\cA}$ and
following three combinations:\\
1) A pseudoautomorphism $\cA\rightarrow\overline{\cA^\star}$. This
transformation is a composition of the pseudoautomorphism
$\cA\rightarrow\overline{\cA}$ with
the automorphism $\cA\rightarrow\cA^\star$.\\
2) A pseudoantiautomorphism\index{pseudoantiautomorphism}
$\cA\rightarrow\overline{\widetilde{\cA}}$. This transformation is a
composition of $\cA\rightarrow\overline{\cA}$ with
the antiautomorphism $\cA\rightarrow\widetilde{\cA}$.\\
3) A pseudoantiautomorphism
$\cA\rightarrow\overline{\widetilde{\cA^\star}}$ (a composition of
$\cA\rightarrow\overline{\cA}$ with the antiautomorphism
$\cA\rightarrow\widetilde{\cA^\star}$).

Thus, we obtain an automorphism set of $\C_n$ consisting of the
eight transformations. Let us show that the set
$\{\Id,\,\star,\,\widetilde{\phantom{cc}},\,\widetilde{\star},\,
\overline{\phantom{cc}},\,\overline{\star},\,
\overline{\widetilde{\phantom{cc}}},\,\overline{\widetilde{\star}}\}$
forms a finite group of order 8 and let us give a physical
interpretation of this group.
\begin{prop}[{\rm\cite{Var04a}}]\label{prop2}\begin{sloppypar}\noindent
Let $\C_n$ be a Clifford algebra over the field $\F=\C$ and let
$\Ext(\C_n)=
\{\Id,\,\star,\,\widetilde{\phantom{cc}},\,\widetilde{\star},\,
\overline{\phantom{cc}},\,\overline{\star},\,
\overline{\widetilde{\phantom{cc}}},\,\overline{\widetilde{\star}}\}$
be an extended automorphism group of the algebra $\C_n$. Then there
is an isomorphism between $\Ext(\C_n)$ and  $CPT/\dZ_2$ group of the
discrete transformations,
$\Ext(\C_n)\simeq\{1,\,P,\,T,\,PT,\,C,\,CP,\,CT,\,CPT\}\simeq
\dZ_2\otimes\dZ_2\otimes\dZ_2$. In this case, space inversion $P$,
time reversal $T$, full reflection $PT$, charge conjugation $C$,
transformations $CP$, $CT$ and the full $CPT$--transformation
correspond to the automorphism $\cA\rightarrow\cA^\star$,
antiautomorphisms $\cA\rightarrow\widetilde{\cA}$,
$\cA\rightarrow\widetilde{\cA^\star}$, pseudoautomorphisms
$\cA\rightarrow\overline{\cA}$,
$\cA\rightarrow\overline{\cA^\star}$, pseudoantiautomorphisms
$\cA\rightarrow\overline{\widetilde{\cA}}$ and
$\cA\rightarrow\overline{\widetilde{\cA^\star}}$,
respectively.\end{sloppypar}
\end{prop}
\begin{proof}\begin{sloppypar}\noindent
The group $\{1,\,P,\,T,\,PT,\,C,\,CP,\,CT,\,CPT\}$ at the conditions
$P^2=T^2=(PT)^2=C^2=(CP)^2=(CT)^2=(CPT)^2=1$ and commutativity of
all the elements forms an Abelian group of order 8, which is
isomorphic to a cyclic group $\dZ_2\otimes\dZ_2\otimes\dZ_2$. The
multiplication table of this group shown in Tab.\,1.\end{sloppypar}
\begin{figure}[ht]
\begin{center}{\renewcommand{\arraystretch}{1.4}
\begin{tabular}{|c||c|c|c|c|c|c|c|c|}\hline
     & $1$  & $P$  & $T$  & $PT$ & $C$  & $CP$ & $CT$ & $CPT$ \\ \hline\hline
$1$  & $1$  & $P$  & $T$  & $PT$ & $C$  & $CP$ & $CT$ & $CPT$ \\
\hline $P$  & $P$  & $1$  & $PT$ & $T$  & $CP$ & $C$  & $CPT$&
$CT$\\ \hline $T$  & $T$  & $PT$ & $1$  & $P$  & $CT$ & $CPT$& $C$
& $CP$\\ \hline $PT$ & $PT$ & $T$  & $P$  & $1$  & $CPT$& $CT$ &
$CP$ & $C$\\ \hline $C$  & $C$  & $CP$ & $CT$ & $CPT$& $1$  & $P$  &
$T$  & $PT$\\ \hline $CP$ & $CP$ & $C$  & $CPT$& $CT$ & $P$  & $1$
& $PT$ & $T$\\ \hline $CT$ & $CT$ & $CPT$& $C$  & $CP$ & $T$  & $PT$
& $1$  & $P$\\ \hline $CPT$& $CPT$& $CT$ & $CP$ & $C$  & $PT$ & $T$
& $P$  & $1$\\ \hline
\end{tabular}
}
\end{center}
\hspace{0.4cm}
\begin{center}{\small \textbf{Tab.\,1:} The multiplication table of the $CPT/\dZ_2$
group.}
\end{center}
\end{figure}
In turn, for the extended automorphism group
$\{\Id,\,\star,\,\widetilde{\phantom{cc}},\,\widetilde{\star},\,
\overline{\phantom{cc}},\,\overline{\star},\,
\overline{\widetilde{\phantom{cc}}},\,\overline{\widetilde{\star}}\}$
in virtue of commutativity $\widetilde{\left(\cA^\star\right)}=
\left(\widetilde{\cA}\right)^\star$,
$\overline{\left(\cA^\star\right)}=\left(\overline{\cA}\right)^\star$,
$\overline{\left(\widetilde{\cA}\right)}=
\widetilde{\left(\overline{\cA}\right)}$,
$\overline{\left(\widetilde{\cA^\star}\right)}=
\widetilde{\left(\overline{\cA}\right)^\star}$ and an involution
property
$\star\star=\widetilde{\phantom{cc}}\widetilde{\phantom{cc}}=
\overline{\phantom{cc}}\;\overline{\phantom{cc}}=\Id$ we have the
multiplication table shown in Tab.\,2.
\begin{figure}[ht]
\begin{center}{\renewcommand{\arraystretch}{1.4}
\begin{tabular}{|c||c|c|c|c|c|c|c|c|}\hline
  & $\Id$ & $\star$ & $\widetilde{\phantom{cc}}$ & $\widetilde{\star}$ &
$\overline{\phantom{cc}}$ & $\overline{\star}$ &
$\overline{\widetilde{\phantom{cc}}}$ &
$\overline{\widetilde{\star}}$ \\ \hline\hline $\Id$ & $\Id$ &
$\star$ & $\widetilde{\phantom{cc}}$ & $\widetilde{\star}$ &
$\overline{\phantom{cc}}$ & $\overline{\star}$ &
$\overline{\widetilde{\phantom{cc}}}$ &
$\overline{\widetilde{\star}}$ \\ \hline $\star$ & $\star$ & $\Id$ &
$\widetilde{\star}$ & $\widetilde{\phantom{cc}}$ &
$\overline{\star}$ & $\overline{\phantom{cc}}$ &
$\overline{\widetilde{\star}}$ &
$\overline{\widetilde{\phantom{cc}}}$\\ \hline
$\widetilde{\phantom{cc}}$ & $\widetilde{\phantom{cc}}$ &
$\overline{\star}$ & $\Id$ & $\star$ &
$\overline{\widetilde{\phantom{cc}}}$ &
$\overline{\widetilde{\star}}$ & $\overline{\phantom{cc}}$ &
$\overline{\star}$\\ \hline $\widetilde{\star}$ &
$\widetilde{\star}$ & $\widetilde{\phantom{cc}}$ & $\star$ & $\Id$ &
$\overline{\widetilde{\star}}$ &
$\overline{\widetilde{\phantom{cc}}}$ & $\overline{\star}$ &
$\overline{\phantom{cc}}$\\ \hline $\overline{\phantom{cc}}$ &
$\overline{\phantom{cc}}$ & $\overline{\star}$ &
$\overline{\widetilde{\phantom{cc}}}$ &
$\overline{\widetilde{\star}}$ & $\Id$ & $\star$ &
$\widetilde{\phantom{cc}}$ & $\widetilde{\star}$\\ \hline
$\overline{\star}$ & $\overline{\star}$ & $\overline{\phantom{cc}}$
& $\overline{\widetilde{\star}}$ &
$\overline{\widetilde{\phantom{cc}}}$ & $\star$ & $\Id$ &
$\widetilde{\star}$ & $\widetilde{\phantom{cc}}$\\ \hline
$\overline{\widetilde{\phantom{cc}}}$ &
$\overline{\widetilde{\phantom{cc}}}$ &
$\overline{\widetilde{\star}}$ & $\overline{\phantom{cc}}$ &
$\overline{\star}$ & $\widetilde{\phantom{cc}}$ &
$\widetilde{\star}$ & $\Id$ & $\star$\\ \hline
$\overline{\widetilde{\star}}$ & $\overline{\widetilde{\star}}$ &
$\overline{\widetilde{\phantom{cc}}}$ & $\overline{\star}$ &
$\overline{\phantom{cc}}$ & $\widetilde{\star}$ &
$\widetilde{\phantom{cc}}$ & $\star$ & $\Id$\\ \hline
\end{tabular}
}
\end{center}
\hspace{0.4cm}
\begin{center}{\small \textbf{Tab.\,2:} The multiplication table of the extended automorphism group.}
\end{center}
\end{figure}
The identity of multiplication tables proves the group isomorphism
\[
\{1,\,P,\,T,\,PT,\,C,\,CP,\,CT,\,CPT\}\simeq
\{\Id,\,\star,\,\widetilde{\phantom{cc}},\,\widetilde{\star},\,
\overline{\phantom{cc}},\,\overline{\star},\,
\overline{\widetilde{\phantom{cc}}},\,\overline{\widetilde{\star}}\}\simeq
\dZ_2\otimes\dZ_2\otimes\dZ_2.
\]
\end{proof}

Further, in the case of $P^2=T^2=\ldots=(CPT)^2=\pm 1$ and
anticommutativity of the elements we have an isomorphism between the
$CPT/\dZ_2$ group and a group $\sExt(\C_n)$. The elements of
$\sExt(\C_n)$ are spinor representations of the automorphisms of the
algebra $\C_n$. As mentioned previously, the Wedderburn-Artin
Theorem allows us to define any spinor representaions for the
automorphisms of $\C_n$. We list these transformations and their
spinor representations (for more details see \cite{Var04a}):
\begin{eqnarray}
\cA\longrightarrow\cA^\star,&&\quad\sA^\star=\sW\sA\sW^{-1},\label{SpinW}\\
\cA\longrightarrow\widetilde{\cA},&&\quad\widetilde{\sA}=\sE\sA^{\sT}\sE^{-1},
\label{SpinE}\\
\cA\longrightarrow\widetilde{\cA^\star},&&\quad\widetilde{\sA^\star}=
\sC\sA^{\sT}\sC^{-1},\quad\sC=\sE\sW,\label{SpinC}\\
\cA\longrightarrow\overline{\cA},&&\quad\overline{\sA}=\Pi\sA^\ast\Pi^{-1},
\label{SpinP}\\
\cA\longrightarrow\overline{\cA^\star},&&\quad\overline{\sA^\star}=
\sK\sA^\ast\sK^{-1},\quad\sK=\Pi\sW,\label{SpinK}\\
\cA\longrightarrow\overline{\widetilde{\cA}},&&\quad
\overline{\widetilde{\sA}}=\sS\left(\sA^{\sT}\right)^\ast\sS^{-1},\quad
\sS=\Pi\sE,\label{SpinS}\\
\cA\longrightarrow\overline{\widetilde{\cA^\star}},&&\quad
\overline{\widetilde{\sA^\star}}=\sF\left(\sA^\ast\right)^{\sT}\sF^{-1},\quad
\sF=\Pi\sC,\label{SpinF}
\end{eqnarray}
where the symbol $\sT$ means a transposition, and $\ast$ is a
complex conjugation. The detailed classification of the extended
automorphism groups $\sExt(\C_n)$ was given in \cite{Var04a}. First
of all, since for the subalgebras $\cl_{p,q}$ over the ring
$\K\simeq\R$ the group $\sExt(\C_n)$ is reduced to $\sAut_\pm(\C_n)$
(reflection group \cite{Var00}), then all the essentially different
groups $\sExt(\C_n)$ correspond to subalgebras $\cl_{p,q}$ with the
quaternionic ring $\K\simeq\BH$, $p-q\equiv 4,6\pmod{8}$. The
classification of the groups $\sExt(\C_n)$ is given with respect to
the subgroups $\sAut_\pm(\cl_{p,q})$. Taking into account the
structure of $\sAut_\pm(\cl_{p,q})$, we have at $p-q\equiv
4,6\pmod{8}$ for the groups
$\sExt(\C_n)=\left\{\sI,\sW,\sE,\sC,\Pi,\sK,\sS,\sF\right\}$ the
following realizations \cite{Var04a}:
\begin{gather}
\sExt^1(\C_n)=\left\{\sI,\cE_{12\cdots p+q},\cE_{j_1j_2\cdots j_k},
\cE_{i_1i_2\cdots i_{p+q-k}},\cE_{\alpha_1\alpha_2\cdots\alpha_a},
\cE_{\beta_1\beta_2\cdots\beta_b},\cE_{c_1c_2\cdots c_s},
\cE_{d_1d_2\cdots d_g}\right\},\nonumber\\
\sExt^2(\C_n)=\left\{\sI,\cE_{12\cdots p+q},\cE_{j_1j_2\cdots j_k},
\cE_{i_1i_2\cdots i_{p+q-k}},\cE_{\beta_1\beta_2\cdots\beta_b},
\cE_{\alpha_1\alpha_2\cdots\alpha_a},\cE_{d_1d_2\cdots d_g},
\cE_{c_1c_2\cdots c_s}\right\},\nonumber\\
\sExt^3(\C_n)=\left\{\sI,\cE_{12\cdots p+q},\cE_{i_1i_2\cdots
i_{p+q-k}}, \cE_{j_1j_2\cdots j_k},
\cE_{\alpha_1\alpha_2\cdots\alpha_a},
\cE_{\beta_1\beta_2\cdots\beta_b},\cE_{d_1d_2\cdots d_g},
\cE_{c_1c_2\cdots c_s}\right\},\nonumber\\
\sExt^4(\C_n)=\left\{\sI,\cE_{12\cdots p+q},\cE_{i_1i_2\cdots
i_{p+q-k}}, \cE_{j_1j_2\cdots j_k},
\cE_{\beta_1\beta_2\cdots\beta_b},
\cE_{\alpha_1\alpha_2\cdots\alpha_a},\cE_{c_1c_2\cdots c_s},
\cE_{d_1d_2\cdots d_g}\right\}.\nonumber
\end{gather}
The groups $\sExt^1(\C_n)$ and $\sExt^2(\C_n)$ have Abelian
subgroups $\sAut_-(\cl_{p,q})$ ($\dZ_2\otimes\dZ_2$ or $\dZ_4$). In
turn, the groups $\sExt^3(\C_n)$ and $\sExt^4(\C_n)$ have
non-Abelian subgroups $\sAut_+(\cl_{p,q})$ ($Q_4/\dZ_2$ or
$D_4/\dZ_2$). The full number of different realizations of
$\sExt(\C_n)$ is 64.

As is known, the Lipschitz group $\Lip_{p,q}$, also called the
Clifford group, introduced by Lipschitz in 1886 \cite{Lips}, may be
defined as the subgroup of invertible elements $s$ of the algebra
$\cl_{p,q}$:
\[
\Lip_{p,q}=\left\{s\in\cl^+_{p,q}\cup\cl^-_{p,q}\;|\;\forall
\bx\in\R^{p,q},\; s\bx s^{-1}\in\R^{p,q}\right\}.
\]
The set $\Lip^+_{p,q}=\Lip_{p,q}\cap\cl^+_{p,q}$ is called {\it
special Lipschitz group}\index{group!special Lipschitz}
\cite{Che55}.

Let $N:\;\cl_{p,q}\rightarrow\cl_{p,q},\;N(\bx)=\bx\widetilde{\bx}$.
If $\bx\in\R^{p,q}$, then $N(\bx)=\bx(-\bx)=-\bx^2=-Q(\bx)$.
Further, the group $\Lip_{p,q}$ has a subgroup
\begin{equation}\label{Pin}
\pin(p,q)=\left\{s\in\Lip_{p,q}\;|\;N(s)=\pm 1\right\}.
\end{equation}
Analogously, {\it a spinor group}\index{group!spinor} $\spin(p,q)$
is defined by the set
\begin{equation}\label{Spin}
\spin(p,q)=\left\{s\in\Lip^+_{p,q}\;|\;N(s)=\pm 1\right\}.
\end{equation}
It is obvious that $\spin(p,q)=\pin(p,q)\cap\cl^+_{p,q}$. The group
$\spin(p,q)$ contains a subgroup
\begin{equation}\label{Spin+}
\spin_+(p,q)=\left\{s\in\spin(p,q)\;|\;N(s)=1\right\}.
\end{equation}
The groups $\GO(p,q),\,\SO(p,q)$ and $\SO_+(p,q)$ are isomorphic,
respectively, to the following quotient groups\index{group!quotient}
\[
\GO(p,q)\simeq\pin(p,q)/\dZ_2,\quad
\SO(p,q)\simeq\spin(p,q)/\dZ_2,\quad
\SO_+(p,q)\simeq\spin_+(p,q)/\dZ_2,
\]
\begin{sloppypar}\noindent
where the kernel\index{kernel} $\dZ_2=\{1,-1\}$. Thus, the groups
$\pin(p,q)$, $\spin(p,q)$ and $\spin_+(p,q)$ are the universal
coverings of the groups $\GO(p,q),\,\SO(p,q)$ and $\SO_+(p,q)$,
respectively.\end{sloppypar}

Over the field $\F=\R$ there exist 64 universal coverings of the
real orthogonal group $\GO(p,q)$:
\[
\rho^{a,b,c,d,e,f,g}:\;\pin^{a,b,c,d,e,f,g}\longrightarrow \GO(p,q),
\]
where
\begin{equation}\label{RCL}
\pin^{a,b,c,d,e,f,g}(p,q)\simeq \frac{(\spin_+(p,q)\odot
C^{a,b,c,d,e,f,g})}{\dZ_2},
\end{equation}
and
\[
C^{a,b,c,d,e,f,g}=\{\pm 1,\,\pm P,\,\pm T,\,\pm PT,\,\pm C,\,\pm
CP,\, \pm CT,\,\pm CPT\}
\]
is {\it a full $CPT$ group} \cite{Var04a,Var04b}.
$C^{a,b,c,d,e,f,g}$ is a finite group of order 16. The group
\[
\Ext(\cl_{p,q})=\frac{C^{a,b,c,d,e,f,g}}{\dZ_2}\simeq CPT/\dZ_2
\]
is called {\it the generating group}. In essence,
$C^{a,b,c,d,e,f,g}$ are five double coverings of the group
$\dZ_2\otimes\dZ_2\otimes\dZ_2$ (extraspecial Salingaros groups, see
\cite{Sal84,Bra85}). All the possible double coverings
$C^{a,b,c,d,e,f,g}$ are given in the Table 3.
\begin{figure}[ht]
\begin{center}{\renewcommand{\arraystretch}{1.2}
\begin{tabular}{|l|l|l|}\hline
$a\;b\;c\;d\;e\;f\;g$ & $C^{a,b,c,d,e,f,g}$ & Type \\ \hline
$+\;+\;+\;+\;+\;+\;+$ & $\dZ_2\otimes\dZ_2\otimes\dZ_2\otimes\dZ_2$ & Abelian \\
three `$+$' and four `$-$' & $\dZ_4\otimes\dZ_2\otimes\dZ_2$ & \\
\hline
one `$+$' and six `$-$' & $Q_4\otimes\dZ_2$ & Non--Abelian \\
five `$+$' and two `$-$' & $D_4\otimes\dZ_2$ &   \\
three `$+$' and four `$-$' &
$\overset{\ast}{\dZ}_4\otimes\dZ_2\otimes\dZ_2$ & \\ \hline
\end{tabular}
}
\end{center}
\hspace{0.3cm}
\begin{center}{\small \textbf{Tab.\,3:} Extraspecial finite groups $C^{a,b,c,d,e,f,g}$ of order 16.}
\end{center}
\end{figure}
The group (\ref{RCL}) with non-Abelian $C^{a,b,c,d,e,f,g}$ is called
{\it Cliffordian group} and respectively {\it non-Cliffordian group}
when $C^{a,b,c,d,e,f,g}$ is Abelian. It is easy to see that in the
case of the algebra $\cl_{p,q}$ (or subalgebra
$\cl_{p,q}\subset\C_n$) with the real division ring $\K\simeq\R$,
$p-q\equiv 0,2\pmod{8}$, $CPT$-structures, defined by the groups
(\ref{RCL}), are reduced to the eight Shirokov-D\c{a}browski
$PT$-structures \cite{Shi58,Shi60,Dab88}.

\section{$CPT$ groups on the representation spaces of $\spin_+(1,3)$}
Let us consider the field
\begin{equation}\label{FieldL}
\boldsymbol{\psi}(\balpha)=\langle x,\fg\,|\boldsymbol{\psi}\rangle,
\end{equation}
where $x\in T_4$, $\fg\in\spin_+(1,3)$. The spinor group
$\spin_+(1,3)\simeq\SU(2)\otimes\SU(2)$ is a universal covering of
the proper orthochronous Lorentz group $\SO_0(1,3)$. The parameters
$x\in T_4$ and $\fg\in\spin_+(1,3)$ describe position and
orientation of the extended object defined by the field
(\ref{FieldL}) (the field on the Poincar\'{e} group). The basic idea
is to define discrete symmetries of the field (\ref{FieldL}) within
the group
\[
\pin^{a,b,c,d,e,f,g}(1,3)\simeq\frac{\spin_+(1,3)\odot
C^{a,b,c,d,e,f,g}}{\dZ_2}.
\]
The automorphisms (discrete symmetries) of
$\pin^{a,b,c,d,e,f,g}(1,3)$ are outer automorphisms with respect to
transformations of the group $\spin_+(1,3)$. We define $CPT$ groups
$C^{a,b,c,d,e,f,g}$ of physical fields of any spin on the
representation spaces of $\spin_+(1,3)$.
\begin{theorem}\label{tinf}
Let
\[
\pin^{a,b,c,d,e,f,g}(1,3)\simeq\frac{\spin_+(1,3)\odot
C^{a,b,c,d,e,f,g}}{\dZ_2}
\]
\begin{sloppypar}\noindent
be the universal covering of the proper Lorentz group $\SO(1,3)$,
where $C^{a,b,c,d,e,f,g}=\{\pm 1, \pm P, \pm T, \pm PT, \pm C, \pm
CP, \pm CT, \pm CPT\}$ is a $CPT$ group of some physical field
defined in the framework of finite-dimensional representation of the
group $\spin_+(1,3)$. At this point, there exits a correspondence
$P\sim\sW$, $T\sim\sE$, $PT\sim\sC$, $C\sim\Pi$, $CP\sim\sK$,
$CT\sim\sS$, $CPT\sim\sF$, where
$\{\sI,\sW,\sE,\sC,\Pi,\sK,\sS,\sF\}\simeq\Ext(\C_n)$ is an
automorphism group of the algebra $\C_n$. Then $CPT$ group of the
field $(l,0)\oplus(0,\dot{l})$ is constructed in the framework of
the finite-dimensional representation
$\fC^{l_0+l_1-1,0}\oplus\fC^{0,l_0-l_1+1}$ of $\spin_+(1,3)$ defined
on the spinspace $\dS_{2^k}\otimes\dS_{2^r}$ with the algebra
\end{sloppypar}
\[
\underbrace{\C_2\otimes\C_2\otimes\cdots\otimes\C_2}_{k\;\text{times}}\bigoplus
\underbrace{\overset{\ast}{\C}_2\otimes\overset{\ast}{\C}_2\otimes\cdots\otimes
\overset{\ast}{\C}_2}_{r\;\text{times}},
\]
where $(l_0,l_1)=\left(\frac{k}{2},\frac{k}{2}+1\right)$,
$(-l_0,l_1)= \left(-\frac{r}{2},\frac{r}{2}+1\right)$. In turn, a
$CPT$ group of the field
$(l^\prime,l^{\prime\prime})\oplus(\dot{l}^{\prime\prime},\dot{l}^\prime)$
is constructed in the framework of representation
$\fC^{l_0+l_1-1,l_0-l_1+1}\oplus\fC^{l_0-l_1+1,l_0+l_1-1}$ of
$\spin_+(1,3)$ defined on the spinspace
$\dS_{2^{k+r}}\oplus\dS_{2^{k+r}}$ with the algebra
\[
\underbrace{\C_2\otimes\C_2\otimes\cdots\otimes\C_2\bigotimes
\overset{\ast}{\C}_2\otimes\overset{\ast}{\C}_2\otimes\cdots\otimes
\overset{\ast}{\C}_2}_{k+r\;\text{times}}\bigoplus
\underbrace{\overset{\ast}{\C}_2\otimes\overset{\ast}{\C}_2\otimes\cdots\otimes
\overset{\ast}{\C}_2\bigotimes\C_2\otimes\C_2\otimes\cdots\otimes\C_2}_{r+k\;\text{times}},
\]
where $(l_0,l_1)=\left(\frac{k-r}{2}, \frac{k+r}{2}+1\right)$.
\end{theorem}
\begin{proof}
As is known, when $\cl_{p,q}$ is simple, then the map
\begin{equation}\label{Simple}
\cl_{p,q}\overset{\gamma}{\longrightarrow}\End_{\K}(\dS),\quad
u\longrightarrow\gamma(u),\quad \gamma(u)\psi=u\psi
\end{equation}\begin{sloppypar}\noindent
gives an irreducible and faithful representation of $\cl_{p,q}$ in
the spinspace $\dS_{2^m}(\K)\simeq I_{p,q}=\cl_{p,q}f$, where
$\psi\in\dS_{2^m}$, $m=\frac{p+q}{2}$.\end{sloppypar}

On the other hand, when $\cl_{p,q}$ is semi-simple, then the map
\begin{equation}\label{Semi-Simple}
\cl_{p,q}\overset{\gamma}{\longrightarrow}\End_{\K\oplus\hat{\K}}
(\dS\oplus\hat{\dS}),\quad u\longrightarrow\gamma(u),\quad
\gamma(u)\psi=u\psi
\end{equation}
gives a faithful but reducible representation of $\cl_{p,q}$ in the
double spinspace\index{spinspace!double} $\dS\oplus\hat{\dS}$, where
$\hat{\dS}=\{\hat{\psi}|\psi\in\dS\}$. In this case, the ideal
$\dS\oplus\hat{\dS}$ possesses a right $\K\oplus\hat{\K}$-linear
structure, $\hat{\K}=\{\hat{\lambda}|\lambda\in\K\}$, and
$\K\oplus\hat{\K}$ is isomorphic to the double division ring
$\R\oplus\R$ when $p-q\equiv 1\pmod{8}$ or to $\BH\oplus\BH$ when
$p-q\equiv 5\pmod{8}$. The map $\gamma$ in (\ref{Simple}) and
(\ref{Semi-Simple}) defines the so called {\it left-regular} spinor
representation\index{representation!spinor} of $\cl(Q)$ in $\dS$ and
$\dS\oplus\hat{\dS}$, respectively. Furthermore, $\gamma$ is {\it
faithful} which means that $\gamma$ is an algebra monomorphism. In
(\ref{Simple}), $\gamma$ is {\it irreducible} which means that $\dS$
possesses no proper (that is, $\neq 0,\,\dS$) invariant
subspaces\index{subspace!invariant} under the left action of
$\gamma(u)$, $u\in\cl_{p,q}$. Representation $\gamma$ in
(\ref{Semi-Simple}) is therefore {\it reducible} since
$\{(\psi,0)|\psi\in\dS\}$ and
$\{(0,\hat{\psi})|\hat{\psi}\in\hat{\dS}\}$ are two proper subspaces
of $\dS\oplus\hat{\dS}$ invariant under $\gamma(u)$ (see
\cite{Lou97,Cru91,Port95}).

Since the spacetime algebra $\cl_{1,3}$ is the simple algebra, then
the map (\ref{Simple}) gives an irreducible representation of
$\cl_{1,3}$ in the spinspace $\dS_2(\BH)$. In turn, representations
of the group $\spin_+(1,3)\in\cl^+_{1,3}\simeq\cl_{3,0}$ are defined
in the spinspace $\dS_2(\C)$.
\begin{sloppypar} Let us consider now spintensor representations of the
group $\fG_+\simeq\SL(2;\C)$ which, as is known, form the base of
all the finite-dimensional representations of the Lorentz group, and
also we consider their relationship with the complex Clifford
algebras. From each complex Clifford algebra
$\C_n=\C\otimes\cl_{p,q}\; (n=p+q)$ we obtain the spinspace
$\dS_{2^{n/2}}$ which is a complexification of the minimal left
ideal of the algebra $\cl_{p,q}$: $\dS_{2^{n/2}}=\C\otimes
I_{p,q}=\C\otimes\cl_{p,q} f_{pq}$, where $f_{pq}$ is the primitive
idempotent of the algebra $\cl_{p,q}$. Further, a spinspace related
with the Pauli algebra $\C_2$ has the form $\dS_2=\C\otimes
I_{2,0}=\C\otimes\cl_{2,0}f_{20}$ or $\dS_2=\C\otimes
I_{1,1}=\C\otimes\cl_{1,1}f_{11}(\C\otimes I_{0,2}=
\C\otimes\cl_{0,2}f_{02})$. Therefore, the tensor product of the $k$
algebras $\C_2$ induces a tensor product of the $k$ spinspaces
$\dS_2$:\end{sloppypar}
%\newpage
\[
\dS_2\otimes\dS_2\otimes\cdots\otimes\dS_2=\dS_{2^k}.
\]
Vectors of the spinspace $\dS_{2^k}$ (or elements of the minimal
left ideal of $\C_{2k}$) are spintensors of the following form:
\begin{equation}\label{6.16}
\boldsymbol{s}^{\alpha_1\alpha_2\cdots\alpha_k}=\sum
\boldsymbol{s}^{\alpha_1}\otimes
\boldsymbol{s}^{\alpha_2}\otimes\cdots\otimes
\boldsymbol{s}^{\alpha_k},
\end{equation}
where summation is produced on all the index collections
$(\alpha_1\ldots\alpha_k)$, $\alpha_i=1,2$. For the each spinor
$\boldsymbol{s}^{\alpha_i}$ from (\ref{6.16}) we have ${}^\prime
\boldsymbol{s}^{\alpha^\prime_i}=
\sigma^{\alpha^\prime_i}_{\alpha_i}\boldsymbol{s}^{\alpha_i}$.
Therefore, in general case we obtain
\begin{equation}\label{6.17}
{}^\prime
\boldsymbol{s}^{\alpha^\prime_1\alpha^\prime_2\cdots\alpha^\prime_k}=\sum
\sigma^{\alpha^\prime_1}_{\alpha_1}\sigma^{\alpha^\prime_2}_{\alpha_2}\cdots
\sigma^{\alpha^\prime_k}_{\alpha_k}\boldsymbol{s}^{\alpha_1\alpha_2\cdots\alpha_k}.
\end{equation}
A representation (\ref{6.17}) is called {\it undotted spintensor
representation of the proper Lorentz group of the rank $k$}.

Further, let $\overset{\ast}{\C}_2$ be the Pauli algebra with the
coefficients which are complex conjugate to the coefficients of
$\C_2$. Let us show that the algebra $\overset{\ast}{\C}_2$ is
derived from $\C_2$ under action of the automorphism
$\cA\rightarrow\cA^\star$ or antiautomorphism
$\cA\rightarrow\widetilde{\cA}$. Indeed, in virtue of an isomorphism
$\C_2\simeq\cl_{3,0}$ a general element
\[
\cA=a^0\e_0+\sum^3_{i=1}a^i\e_i+\sum^3_{i=1}\sum^3_{j=1}a^{ij}\e_{ij}+
a^{123}\e_{123}
\]
of the algebra $\cl_{3,0}$ can be written in the form
\begin{equation}\label{6.17'}
\cA=(a^0+\omega a^{123})\e_0+(a^1+\omega a^{23})\e_1+(a^2+\omega
a^{31})\e_2 +(a^3+\omega a^{12})\e_3,
\end{equation}
where $\omega=\e_{123}$. Since $\omega$ belongs to a center of the
algebra $\cl_{3,0}$ ($\omega$ commutes with all the basis elements)
and $\omega^2=-1$, then we can to suppose $\omega\equiv i$. The
action of the automorphism $\star$ on the homogeneous element $\cA$
of the degree $k$ is defined by the formula $\cA^\star=(-1)^k\cA$.
In accordance with this the action of the automorphism
$\cA\rightarrow\cA^\star$, where $\cA$ is the element (\ref{6.17'}),
has the form
\begin{equation}\label{In1}
\cA\longrightarrow\cA^\star=-(a^0-\omega a^{123})\e_0-(a^1-\omega
a^{23})\e_1 -(a^2-\omega a^{31})\e_2-(a^3-\omega a^{12})\e_3.
\end{equation}
Therefore, $\star:\,\C_2\rightarrow -\overset{\ast}{\C}_2$.
Correspondingly, the action of the antiautomorphism
$\cA\rightarrow\widetilde{\cA}$ on the homogeneous element $\cA$ of
the degree $k$ is defined by the formula
$\widetilde{\cA}=(-1)^{\frac{k(k-1)}{2}}\cA$. Thus, for the element
(\ref{6.17'}) we obtain
\begin{equation}\label{In2}
\cA\longrightarrow\widetilde{\cA}=(a^0-\omega a^{123})\e_0+
(a^1-\omega a^{23})\e_1+(a^2-\omega a^{31})\e_2+(a^3-\omega
a^{12})\e_3,
\end{equation}
that is, $\widetilde{\phantom{cc}}:\,\C_2\rightarrow
\overset{\ast}{\C}_2$. This allows us to define an algebraic
analogue of the Wigner's representation doubling:
$\C_2\oplus\overset{\ast}{\C}_2$. Further, from (\ref{6.17'}) it
follows that
$\cA=\cA_1+\omega\cA_2=(a^0\e_0+a^1\e_1+a^2\e_2+a^3\e_3)+
\omega(a^{123}\e_0+a^{23}\e_1+a^{31}\e_2+a^{12}\e_3)$. In general
case, by virtue of an isomorphism $\C_{2k}\simeq\cl_{p,q}$, where
$\cl_{p,q}$ is a real Clifford algebra with a division ring
$\K\simeq\C$, $p-q\equiv 3,7 \pmod{8}$, we have for the general
element of $\cl_{p,q}$ an expression $\cA=\cA_1+\omega\cA_2$, here
$\omega^2=\e^2_{12\ldots p+q}=-1$ and, therefore, $\omega\equiv i$.
Thus, from $\C_{2k}$ under action of the automorphism
$\cA\rightarrow\cA^\star$ we obtain a general algebraic doubling
\begin{equation}\label{D}
\C_{2k}\oplus\overset{\ast}{\C}_{2k}.
\end{equation}

The tensor product
$\overset{\ast}{\C}_2\otimes\overset{\ast}{\C}_2\otimes\cdots\otimes
\overset{\ast}{\C}_2\simeq\overset{\ast}{\C}_{2r}$ of the $r$
algebras $\overset{\ast}{\C}_2$ induces the tensor product of the
$r$ spinspaces $\dot{\dS}_2$:
\[
\dot{\dS}_2\otimes\dot{\dS}_2\otimes\cdots\otimes\dot{\dS}_2=\dot{\dS}_{2^r}.
\]
Vectors of the spinspace $\dot{\dS}_{2^r}$ has the form
\begin{equation}\label{6.18}
\boldsymbol{s}^{\dot{\alpha}_1\dot{\alpha}_2\cdots\dot{\alpha}_r}=\sum
\boldsymbol{s}^{\dot{\alpha}_1}\otimes
\boldsymbol{s}^{\dot{\alpha}_2}\otimes\cdots\otimes
\boldsymbol{s}^{\dot{\alpha}_r},
\end{equation}
where the each cospinor $\boldsymbol{s}^{\dot{\alpha}_i}$ from
(\ref{6.18}) is transformed by the rule ${}^\prime
\boldsymbol{s}^{\dot{\alpha}^\prime_i}=
\sigma^{\dot{\alpha}^\prime_i}_{\dot{\alpha}_i}\boldsymbol{s}^{\dot{\alpha}_i}$.
Therefore,
\begin{equation}\label{6.19}
{}^\prime
\boldsymbol{s}^{\dot{\alpha}^\prime_1\dot{\alpha}^\prime_2\cdots
\dot{\alpha}^\prime_r}=\sum\sigma^{\dot{\alpha}^\prime_1}_{\dot{\alpha}_1}
\sigma^{\dot{\alpha}^\prime_2}_{\dot{\alpha}_2}\cdots
\sigma^{\dot{\alpha}^\prime_r}_{\dot{\alpha}_r}
\boldsymbol{s}^{\dot{\alpha}_1\dot{\alpha}_2\cdots\dot{\alpha}_r}.
\end{equation}\begin{sloppypar}\noindent
The representation (\ref{6.19}) is called {\it a dotted spintensor
representation of the proper Lorentz group of the rank
$r$}.\end{sloppypar}

In general case we have a tensor product of the $k$ algebras $\C_2$
and the $r$ algebras $\overset{\ast}{\C}_2$:
\[
\C_2\otimes\C_2\otimes\cdots\otimes\C_2\bigotimes
\overset{\ast}{\C}_2\otimes
\overset{\ast}{\C}_2\otimes\cdots\otimes\overset{\ast}{\C}_2\simeq
\C_{2k}\otimes\overset{\ast}{\C}_{2r},
\]
which induces a spinspace
\[
\dS_2\otimes\dS_2\otimes\cdots\otimes\dS_2\bigotimes\dot{\dS}_2\otimes
\dot{\dS}_2\otimes\cdots\otimes\dot{\dS}_2=\dS_{2^{k+r}}
\]
with the vectors
\begin{equation}\label{6.20'}
\boldsymbol{s}^{\alpha_1\alpha_2\cdots\alpha_k\dot{\alpha}_1\dot{\alpha}_2\cdots
\dot{\alpha}_r}=\sum \boldsymbol{s}^{\alpha_1}\otimes
\boldsymbol{s}^{\alpha_2}\otimes\cdots\otimes
\boldsymbol{s}^{\alpha_k}\otimes
\boldsymbol{s}^{\dot{\alpha}_1}\otimes
\boldsymbol{s}^{\dot{\alpha}_2}\otimes\cdots\otimes
\boldsymbol{s}^{\dot{\alpha}_r}.
\end{equation}
In this case we have a natural unification of the representations
(\ref{6.17}) and (\ref{6.19}):
\begin{equation}\label{6.20}
{}^\prime
\boldsymbol{s}^{\alpha^\prime_1\alpha^\prime_2\cdots\alpha^\prime_k
\dot{\alpha}^\prime_1\dot{\alpha}^\prime_2\cdots\dot{\alpha}^\prime_r}=\sum
\sigma^{\alpha^\prime_1}_{\alpha_1}\sigma^{\alpha^\prime_2}_{\alpha_2}\cdots
\sigma^{\alpha^\prime_k}_{\alpha_k}\sigma^{\dot{\alpha}^\prime_1}_{
\dot{\alpha}_1}\sigma^{\dot{\alpha}^\prime_2}_{\dot{\alpha}_2}\cdots
\sigma^{\dot{\alpha}^\prime_r}_{\dot{\alpha}_r}
\boldsymbol{s}^{\alpha_1\alpha_2\cdots\alpha_k\dot{\alpha}_1\dot{\alpha}_2\cdots
\dot{\alpha}_r}.
\end{equation}
So, a representation (\ref{6.20}) is called {\it a spintensor
representation of the proper Lorentz group of the rank $(k,r)$}.

Further, let $\fg\rightarrow T_{\fg}$ be an arbitrary linear
representation of the proper orthochronous Lorentz group
$\fG_+=\SO_0(1,3)$ and let $\sA_i(t)=T_{a_i(t)}$ be an infinitesimal
operator corresponding to the rotation $a_i(t)\in\fG_+$.
Analogously, let $\sB_i(t)=T_{b_i(t)}$, where $b_i(t)\in\fG_+$ is
the hyperbolic rotation. The operators $\sA_i$ and $\sB_i$ satisfy
to the following relations:
\begin{equation}\label{Com1}
\left.\begin{array}{lll} \ld\sA_1,\sA_2\rd=\sA_3, &
\ld\sA_2,\sA_3\rd=\sA_1, &
\ld\sA_3,\sA_1\rd=\sA_2,\\[0.1cm]
\ld\sB_1,\sB_2\rd=-\sA_3, & \ld\sB_2,\sB_3\rd=-\sA_1, &
\ld\sB_3,\sB_1\rd=-\sA_2,\\[0.1cm]
\ld\sA_1,\sB_1\rd=0, & \ld\sA_2,\sB_2\rd=0, &
\ld\sA_3,\sB_3\rd=0,\\[0.1cm]
\ld\sA_1,\sB_2\rd=\sB_3, & \ld\sA_1,\sB_3\rd=-\sB_2, & \\[0.1cm]
\ld\sA_2,\sB_3\rd=\sB_1, & \ld\sA_2,\sB_1\rd=-\sB_3, & \\[0.1cm]
\ld\sA_3,\sB_1\rd=\sB_2, & \ld\sA_3,\sB_2\rd=-\sB_1. &
\end{array}\right\}
\end{equation}
Denoting $\sI^{23}=\sA_1$, $\sI^{31}=\sA_2$, $\sI^{12}=\sA_3$, and
$\sI^{01}=\sB_1$, $\sI^{02}=\sB_2$, $\sI^{03}=\sB_3$ we write the
relations (\ref{Com1}) in a more compact form:
\[
\ld\sI^{\mu\nu},\sI^{\lambda\rho}\rd=\delta_{\mu\rho}\sI^{\lambda\nu}+
\delta_{\nu\lambda}\sI^{\mu\rho}-\delta_{\nu\rho}\sI^{\mu\lambda}-
\delta_{\mu\lambda}\sI^{\nu\rho}.
\]

As is known \cite{GMS}, finite-dimensional (spinor) representations
of the group $\SO_0(1,3)$ in the space of symmetrical polynomials
$\Sym_{(k,r)}$ have the following form:
\begin{equation}\label{TenRep}
T_{\fg}q(\xi,\overline{\xi})=(\gamma\xi+\delta)^{l_0+l_1-1}
\overline{(\gamma\xi+\delta)}^{l_0-l_1+1}q\left(\frac{\alpha\xi+\beta}{\gamma\xi+\delta};
\frac{\overline{\alpha\xi+\beta}}{\overline{\gamma\xi+\delta}}\right),
\end{equation}
where $k=l_0+l_1-1$, $r=l_0-l_1+1$, and the pair $(l_0,l_1)$ defines
some representation of the group $\SO_0(1,3)$ in the
Gel'fand-Naimark basis:
\[
H_{3}\xi_{k\nu} =m\xi_{k\nu},
\]
\[
H_{+}\xi_{k\nu} =\sqrt{(k+\nu+1)(k-\nu)}\xi_{k,\nu+1},
\]
\[
H_{-}\xi_{k\nu} =\sqrt{(k+\nu)(k-\nu+1)}\xi_{k,\nu-1},
\]
\[
F_{3}\xi_{k\nu}
=C_{l}\sqrt{k^{2}-\nu^{2}}\xi_{k-1,\nu}-A_{l}\nu\xi_{k,\nu}-C_{k+1}\sqrt{(k+1)^{2}-\nu^{2}}\xi_{k+1,\nu},
\]
\begin{multline}
F_{+}\xi_{k\nu} =C_{k}\sqrt{(k-\nu)(k-\nu-1)}\xi_{k-1,\nu+1}
-A_{k}\sqrt{(k-\nu)(k+\nu+1)}\xi_{k,\nu+1}+ \\
+C_{k+1}\sqrt{(k+\nu+1)(k+\nu+2)}\xi_{k+1,\nu+1},\nonumber
\end{multline}
\begin{multline}
F_{-}\xi_{k\nu} =-C_{k}\sqrt{(k+\nu)(k+\nu-1)}\xi_{k-1,\nu-1}-A_{k}\sqrt{(k+\nu)(k-\nu+1)}\xi_{k,\nu-1}-\\
-C_{k+1}\sqrt{(k-\nu+1)(k-\nu+2)}\xi_{k+1,\nu-1},\nonumber
\end{multline}
\begin{equation}\label{GNB}
A_{k}=\frac{\bi l_{0}l_{1}}{k(k+1)},\quad
C_{k}=\frac{\bi}{k}\sqrt{\frac{(k^{2}-l^{2}_{0})(k^{2}-l^{2}_{1})}
{4k^{2}-1}},
\end{equation}
$$\nu=-k,-k+1,\ldots,k-1,k,$$
$$k=l_{0}\,,l_{0}+1,\ldots,$$
where $l_{0}$ is positive integer or half-integer number, $l_{1}$ is
an arbitrary complex number. These formulae define a
finite--dimensional representation of the group $\SO_0(1,3)$ when
$l^2_1=(l_0+p)^2$, $p$ is some natural number. In the case
$l^2_1\neq(l_0+p)^2$ we have an infinite-dimensional representation
of $\SO_0(1,3)$. The operators $H_{3},H_{+},H_{-},F_{3},F_{+},F_{-}$
are
\begin{eqnarray}
&&H_+=\bi\sA_1-\sA_2,\quad H_-=\bi\sA_1+\sA_2,\quad H_3=\bi\sA_3,\nonumber\\
&&F_+=\bi\sB_1-\sB_2,\quad F_-=\bi\sB_1+\sB_2,\quad
F_3=\bi\sB_3.\nonumber
\end{eqnarray}
Let us consider the operators
\begin{gather}
\sX_l=\frac{1}{2}\bi(\sA_l+\bi\sB_l),\quad\sY_l=\frac{1}{2}\bi(\sA_l-\bi\sB_l),
\label{SL25}\\
(l=1,2,3).\nonumber
\end{gather}
Using the relations (\ref{Com1}), we find that
\begin{equation}\label{Com2}
\ld\sX_k,\sX_l\rd=\bi\varepsilon_{klm}\sX_m,\quad
\ld\sY_l,\sY_m\rd=\bi\varepsilon_{lmn}\sY_n,\quad
\ld\sX_l,\sY_m\rd=0.
\end{equation}
Further, introducing generators of the form
\begin{equation}\label{SL26}
\left.\begin{array}{cc}
\sX_+=\sX_1+\bi\sX_2, & \sX_-=\sX_1-\bi\sX_2,\\[0.1cm]
\sY_+=\sY_1+\bi\sY_2, & \sY_-=\sY_1-\bi\sY_2,
\end{array}\right\}
\end{equation}
we see that in virtue of commutativity of the relations (\ref{Com2})
a space of an irreducible finite--dimensional representation of the
group $\SL(2,\C)$ can be spanned on the totality of
$(2l+1)(2\dot{l}+1)$ basis vectors $\mid
l,m;\dot{l},\dot{m}\rangle$, where $l,m,\dot{l},\dot{m}$ are integer
or half--integer numbers, $-l\leq m\leq l$, $-\dot{l}\leq
\dot{m}\leq \dot{l}$. Therefore,
\begin{eqnarray}
&&\sX_-\mid l,m;\dot{l},\dot{m}\rangle= \sqrt{(l+m)(l-m+1)}\mid
l,m-1,\dot{l},\dot{m}\rangle
\;\;(m>-l),\nonumber\\
&&\sX_+\mid l,m;\dot{l},\dot{m}\rangle= \sqrt{(l-m)(l+m+1)}\mid
l,m+1;\dot{l},\dot{m}\rangle
\;\;(m<l),\nonumber\\
&&\sX_3\mid l,m;\dot{l},\dot{m}\rangle=
m\mid l,m;\dot{l},\dot{m}\rangle,\nonumber\\
&&\sY_-\mid l,m;\dot{l},\dot{m}\rangle=
\sqrt{(\dot{l}+\dot{m})(\dot{l}-\dot{m}+1)}\mid
l,m;\dot{l},\dot{m}-1
\rangle\;\;(\dot{m}>-\dot{l}),\nonumber\\
&&\sY_+\mid l,m;\dot{l},\dot{m}\rangle=
\sqrt{(\dot{l}-\dot{m})(\dot{l}+\dot{m}+1)}\mid
l,m;\dot{l},\dot{m}+1
\rangle\;\;(\dot{m}<\dot{l}),\nonumber\\
&&\sY_3\mid l,m;\dot{l},\dot{m}\rangle= \dot{m}\mid
l,m;\dot{l},\dot{m}\rangle.\label{Waerden}
\end{eqnarray}
From the relations (\ref{Com2}) it follows that each of the sets of
infinitesimal operators $\sX$ and $\sY$ generates the group $\SU(2)$
and these two groups commute with each other. Thus, from the
relations (\ref{Com2}) and (\ref{Waerden}) it follows that the group
$\SL(2,\C)$, in essence, is equivalent locally to the group
$\SU(2)\otimes\SU(2)$. In contrast to the Gel'fand--Naimark
representation for the Lorentz group \cite{GMS,Nai58}, which does
not find a broad application in physics, a representation
(\ref{Waerden}) is a most useful in theoretical physics (see, for
example, \cite{AB,Sch61,RF,Ryd85}). This representation for the
Lorentz group was first given by Van der Waerden in \cite{Wa32}. It
should be noted here that the representation basis, defined by the
formulae (\ref{SL25})--(\ref{Waerden}), has an evident physical
meaning. For example, in the case of
$(1,0)\oplus(0,1)$--representation space there is an analogy with
the photon spin states. Namely, the operators $\sX$ and $\sY$
correspond to the right and left polarization states of the photon.
The following relations between generators $\sY_\pm$, $\sX_\pm$,
$\sY_3$, $\sX_3$ and $H_\pm$, $F_\pm$, $H_3$, $F_3$ define a
relationship between the Van der Waerden and Gel'fand-Naimark bases:
\[
{\renewcommand{\arraystretch}{1.7}
\begin{array}{ccc}
\sY_+&=&-\dfrac{1}{2}(F_++\bi H_+),\\
\sY_-&=&-\dfrac{1}{2}(F_-+\bi H_-),\\
\sY_3&=&-\dfrac{1}{2}(F_3+\bi H_3),
\end{array}\quad
\begin{array}{ccc}
\sX_+&=&\dfrac{1}{2}(F_+-\bi H_+),\\
\sX_-&=&\dfrac{1}{2}(F_--\bi H_-),\\
\sX_3&=&\dfrac{1}{2}(F_3-\bi H_3).
\end{array}
}
\]
The relation between the numbers $l_0$, $l_1$ and the number $l$
(the weight of representation in the basis (\ref{Waerden})) is given
by the following formula:
\[
(l_0,l_1)=\left(l,l+1\right).
\]
Whence it immediately follows that
\begin{equation}\label{RelL}
l=\frac{l_0+l_1-1}{2}.
\end{equation}
As is known \cite{GMS}, if an irreducible representation of the
proper Lorentz group $\SO_0(1,3)$ is defined by the pair
$(l_0,l_1)$, then a conjugated representation is also irreducible
and is defined by a pair $\pm(l_0,-l_1)$. Therefore,
\[
(l_0,l_1)=\left(-\dot{l},\,\dot{l}+1\right).
\]
Thus,
\begin{equation}\label{RelDL}
\dot{l}=\frac{l_0-l_1+1}{2}.
\end{equation}

Further, representations $\boldsymbol{\tau}_{s_1,s_2}$ and
$\boldsymbol{\tau}_{s^\prime_1,s^\prime_2}$ are called
\emph{interlocking irreducible representations of the Lorentz group
}, that is, such representations that
$s^\prime_1=s_1\pm\frac{1}{2}$, $s^\prime_2=s_2\pm\frac{1}{2}$
\cite{GY48}. The two most full schemes of the interlocking
irreducible representations of the Lorentz group (Gel'fand-Yaglom
chains) for integer and half-integer spins are shown on the Fig.\,1
and Fig.\,2.
\begin{figure}[ht]
\[
\dgARROWLENGTH=0.5em \dgHORIZPAD=1.7em \dgVERTPAD=2.2ex
\begin{diagram}
\node[5]{(s,0)}\arrow{e,-}\arrow{s,-}\node{\cdots}\\
\node[5]{\vdots}\arrow{s,-}\\
\node[3]{(2,0)}\arrow{e,-}\arrow{s,-}\node{\cdots}\arrow{e,-}
\node{\left(\frac{s+2}{2},\frac{s-2}{2}\right)}\arrow{s,-}\arrow{e,-}
\node{\cdots}\\
\node[2]{(1,0)}\arrow{s,-}\arrow{e,-}
\node{\left(\frac{3}{2},\frac{1}{2}\right)}\arrow{s,-}\arrow{e,-}
\node{\cdots}\arrow{e,-}
\node{\left(\frac{s+1}{2},\frac{s-1}{2}\right)}\arrow{s,-}\arrow{e,-}
\node{\cdots}\\
\node{(0,0)}\arrow{e,-}
\node{\left(\frac{1}{2},\frac{1}{2}\right)}\arrow{s,-}\arrow{e,-}
\node{(1,1)}\arrow{s,-}\arrow{e,-}\node{\cdots}\arrow{e,-}
\node{\left(\frac{s}{2},\frac{s}{2}\right)}\arrow{s,-}\arrow{e,-}
\node{\cdots}\\
\node[2]{(0,1)}\arrow{e,-}
\node{\left(\frac{1}{2},\frac{3}{2}\right)}\arrow{s,-}\arrow{e,-}
\node{\cdots}\arrow{e,-}
\node{\left(\frac{s-1}{2},\frac{s+1}{2}\right)}\arrow{s,-}\arrow{e,-}
\node{\cdots}\\
\node[3]{(0,2)}\arrow{e,-}\node{\cdots}\arrow{e,-}
\node{\left(\frac{s-2}{2},\frac{s+2}{2}\right)}\arrow{s,-}\arrow{e,-}
\node{\cdots}\\
\node[5]{\vdots}\arrow{s,-}\\
\node[5]{(0,s)}\arrow{e,-}\node{\cdots}
\end{diagram}
\]
\begin{center}{\small {\bf Fig.\,1:} Interlocking representation scheme for the fields of integer spin
(Bose-scheme).}\end{center}
\end{figure}
\begin{figure}[ht]
\[
\dgARROWLENGTH=0.5em
\dgHORIZPAD=1.7em %1.5em
\dgVERTPAD=2.2ex %2ex
\begin{diagram}
\node[4]{(s,0)}\arrow{s,-}\arrow{e,-}\node{\cdots}\\
\node[4]{\vdots}\arrow{s,-}\\
\node[2]{\left(\frac{3}{2},0\right)}\arrow{s,-}\arrow{e,-}
\node{\cdots}\arrow{e,-}
\node{\left(\frac{2s+3}{4},\frac{2s-3}{4}\right)}\arrow{s,-}\arrow{e,-}
\node{\cdots}\\
\node{\left(\frac{1}{2},0\right)}\arrow{s,-}\arrow{e,-}
\node{\left(1,\frac{1}{2}\right)}\arrow{s,-}\arrow{e,-}
\node{\cdots}\arrow{e,-}
\node{\left(\frac{2s+1}{4},\frac{2s-1}{4}\right)}\arrow{s,-}\arrow{e,-}
\node{\cdots}\\
\node{\left(0,\frac{1}{2}\right)}\arrow{e,-}
\node{\left(\frac{1}{2},1\right)}\arrow{s,-}\arrow{e,-}
\node{\cdots}\arrow{e,-}
\node{\left(\frac{2s-1}{4},\frac{2s+1}{4}\right)}\arrow{s,-}\arrow{e,-}
\node{\cdots}\\
\node[2]{\left(0,\frac{3}{2}\right)}\arrow{e,-}
\node{\cdots}\arrow{e,-}
\node{\left(\frac{2s-3}{4},\frac{2s+3}{4}\right)}\arrow{s,-}\arrow{e,-}
\node{\cdots}\\
\node[4]{\vdots}\arrow{s,-}\\
\node[4]{(0,s)}\arrow{e,-}\node{\cdots}
\end{diagram}
\]
\begin{center}{\small {\bf Fig.\,2:} Interlocking representation scheme for the fields of half-integer spin
(Fermi-scheme).}\end{center}
\end{figure}
As follows from Fig.\,1 the simplest field is the scalar field
\[
(0,0).
\]
This field is described by the Fock-Klein-Gordon equation. In its
turn, the simplest field from the Fermi-scheme (Fig.\,2) is the
electron-positron (spinor) field corresponding to the following
interlocking scheme:
\[
\dgARROWLENGTH=2.5em
\begin{diagram}
\node{\left(\frac{1}{2},0\right)}\arrow{e,<>}
\node{\left(0,\frac{1}{2}\right)}
\end{diagram}.
\]
\begin{sloppypar}\noindent
This field is described by the Dirac equation. Further, the next
field from the Bose-scheme (Fig.\,1) is a photon field (Maxwell
field) defined within the interlocking scheme\end{sloppypar}
\[
\dgARROWLENGTH=2.5em
\begin{diagram}
\node{(1,0)}\arrow{e,<>}\node{\left(\frac{1}{2},\frac{1}{2}\right)}
\arrow{e,<>}\node{(0,1)}
\end{diagram}.
\]
This interlocking scheme leads to the Maxwell equations. The fields
$(1/2,0)\oplus(0,1/2)$ and $(1,0)\oplus(0,1)$ (Dirac and Maxwell
fields) are particular cases of fields of the type
$(l,0)\oplus(0,l)$. Wave equations for such fields and their general
solutions were found in the works \cite{Var03,Var04e,Var05b}.

It is easy to see that the interlocking scheme, corresponded to the
Maxwell field, contains the field of tensor type:
\[
\left(\frac{1}{2},\frac{1}{2}\right).
\]
Further, the next interlocking scheme (see Fig.\,2)
\[
\dgARROWLENGTH=2.5em
\begin{diagram}
\node{\left(\frac{3}{2},0\right)}\arrow{e,<>}\node{\left(1,\frac{1}{2}\right)}
\arrow{e,<>}\node{\left(\frac{1}{2},1\right)}\arrow{e,<>}\node{\left(0,\frac{3}{2}\right)}
\end{diagram},
\]
corresponding to the Pauli-Fierz equations \cite{FP39}, contains a
chain of the type
\[
\dgARROWLENGTH=2.5em
\begin{diagram}
\node{\left(1,\frac{1}{2}\right)}\arrow{e,<>}
\node{\left(\frac{1}{2},1\right)}
\end{diagram}.
\]
In such a way we come to wave equations for the fields
$\boldsymbol{\psi}(\balpha)=\langle
x,\fg\,|\boldsymbol{\psi}\rangle$ of tensor type
$(l_1,l_2)\oplus(l_2,l_1)$. Wave equations for such fields and their
general solutions were found in the work \cite{Var07b}.

A relation between the numbers $l_0$, $l_1$ of the Gel'fand-Naimark
representation (\ref{GNB}) and the number $k$ of the factors $\C_2$
in the product $\C_2\otimes\C_2\otimes\cdots\otimes\C_2$ is given by
the following formula:
\[
(l_0,l_1)=\left(\frac{k}{2},\frac{k}{2}+1\right),
\]
Hence it immediately follows that $k=l_0+l_1-1$. Thus, we have {\it
a complex representation $\fC^{l_0+l_1-1,0}$  of the group
$\spin_+(1,3)$ in the spinspace $\dS_{2^k}$}. If the representation
$\fC^{l_0+l_1-1,0}$ is reducible, then the space $\dS_{2^{k}}$ is
decomposed into a direct sum of irreducible subspaces, that is, it
is possible to choose in $\dS_{2^{k}}$ such a basis, in which all
the matrices take a block-diagonal form. Then the field
$\boldsymbol{\psi}(\balpha)$ is reduced to some number of the fields
corresponding to irreducible representations of the group
$\spin_+(1,3)$, each of which is transformed independently from the
other, and the field $\boldsymbol{\psi}(\balpha)$ in this case is a
collection of the fields with more simple structure. It is obvious
that these more simple fields correspond to irreducible
representations $\fC$.

Analogously, a relation between the numbers $l_0$, $l_1$ of the
Gel'fand-Naimark representation (\ref{GNB}) and the number $r$ of
the factors $\overset{\ast}{\C}_2$ in the product
$\overset{\ast}{\C}_2\otimes
\overset{\ast}{\C}_2\otimes\cdots\otimes\overset{\ast}{\C}_2$ is
given by the following formula:
\[
(-l_0,l_1)=\left(-\frac{r}{2},\frac{r}{2}+1\right).
\]
Hence it immediately follows that $r=l_0-l_1+1$. Thus, we have a
complex representation $\fC^{0,l_0-l_1+1}$ of $\spin_+(1,3)$ in the
spinspace $\dS_{2^r}$.

As is known \cite{Nai58,GMS,RF}, a system of irreducible
finite-dimensional representations of the group $\fG_+$ is realized
in the space $\Sym_{(k,r)}\subset\dS_{2^{k+r}}$ of symmetric
spintensors. The dimensionality of $\Sym_{(k,r)}$ is equal to
$(k+1)(r+1)$. A representation of the group $\fG_+$, defined by such
spintensors, is irreducible and denoted by the symbol
$\fD^{(l,\dot{l})}(\sigma)$, where $2l=k,\;2\dot{l}=r$, the numbers
$l$ and $\dot{l}$ are integer or half-integer. In general case, the
field $\boldsymbol{\psi}(\balpha)$ is the field of type
$(l,\dot{l})$. As a rule, in physics there are two basic types of
the fields:\\[0.3cm]
1) The field of type $(l,0)$. The structure of this field (or the
field $(0,\dot{l})$) is described by the representation
$\fD^{(l,0)}(\sigma)$ ($\fD^{(0,\dot{l})}(\sigma)$), which is
realized in the space $\dS_{2^k}$ ($\dS_{2^r}$). At this point, the
algebra $\C_{2k}\simeq\C_2\otimes\C_2\otimes\cdots\otimes\C_2$
(correspondingly,
$\overset{\ast}{\C}_{2k}\simeq\overset{\ast}{\C}_2\otimes
\overset{\ast}{\C}_2\otimes\cdots\otimes\overset{\ast}{\C}_2$) is
associated with the field of the type $(l,0)$ (correspondingly,
$(0,\dot{l})$). The trivial case $l=0$ corresponds to {\it a
Pauli-Weisskopf field} describing the scalar particles. Further, at
$l=\dot{l}=1/2$ we have {\it a Weyl field} describing the neutrino.
At this point the antineutrino is described by a fundamental
representation $\fD^{(1/2,0)}(\sigma)=\sigma$ of the group $\fG_+$
and the algebra $\C_2$. Correspondingly, the neutrino is described
by a conjugated representation $\fD^{(0,1/2)}(\sigma)$ and the
algebra $\overset{\ast}{\C}_2$. In essence, one can say that the
algebra $\C_2$ ($\overset{\ast}{\C}_2$) is the basic building block,
from which other physical fields
built by means of direct sum or tensor product. One can say that this situation looks like
the de Broglie fusion method \cite{Bro43}\\[0.3cm]
2) The field of type $(l,0)\oplus(0,\dot{l})$. The structure of this
field admits a space inversion and, therefore, in accordance with a
Wigner's doubling \cite{Wig64} is described by a representation
$\fD^{(l,0)}\oplus \fD^{(0,\dot{l})}$ of the group $\fG_+$. This
representation is realized in the space $\dS_{2^{2k}}$. The Clifford
algebra, related with this representation, is a direct sum
$\C_{2k}\oplus\overset{\ast}{\C}_{2k}\simeq
\C_2\otimes\C_2\otimes\cdots\otimes\C_2\bigoplus\overset{\ast}{\C}_2\otimes
\overset{\ast}{\C}_2\otimes\cdots\otimes\overset{\ast}{\C}_2$. In
the simplest case $l=1/2$ we have {\it bispinor (electron--positron)
Dirac field} $(1/2,0)\oplus(0,1/2)$ with the algebra $\C_2\oplus
\overset{\ast}{\C}_2$. It should be noted that the Dirac algebra
$\C_4$, considered as a tensor product $\C_2\otimes\C_2$ (or
$\C_2\otimes\overset{\ast}{\C}_2$) in accordance with (\ref{6.16})
(or (\ref{6.20'})) gives rise to spintensors
$\boldsymbol{s}^{\alpha_1\alpha_2}$ (or
$\boldsymbol{s}^{\alpha_1\dot{\alpha}_1}$), but it contradicts with
the usual definition of the Dirac bispinor as a pair
$(\boldsymbol{s}^{\alpha_1},\boldsymbol{s}^{\dot{\alpha}_1})$.
Therefore, the Clifford algebra, associated with the Dirac field, is
$\C_2\oplus\overset{\ast}{\C}_2$, and a spinspace of this sum in
virtue of unique decomposition $\dS_2\oplus\dot{\dS}_2=\dS_4$ is a
spinspace of $\C_4$.

Spinor representations of the units of $\C_n$ we will define in the
Brauer-Weyl representation \cite{BW35}:
\begin{equation}\label{BWR}
{\renewcommand{\arraystretch}{1.2}
\begin{array}{lcl}
\cE_{1}&=&\sigma_{1}\otimes\boldsymbol{1}_2\otimes\cdots\otimes
\boldsymbol{1}_2\otimes
\boldsymbol{1}_2\otimes\boldsymbol{1}_2,\\
\cE_{2}&=&\sigma_{3}\otimes\sigma_{1}\otimes\boldsymbol{1}_2\otimes\cdots\otimes
\boldsymbol{1}_2\otimes\boldsymbol{1}_2,\\
\cE_{3}&=&\sigma_{3}\otimes\sigma_{3}\otimes\sigma_{1}\otimes\boldsymbol{1}_2
\otimes\cdots\otimes\boldsymbol{1}_2,\\
\hdotsfor[2]{3}\\
\cE_{m}&=&\sigma_{3}\otimes\sigma_{3}\otimes\cdots\otimes\sigma_{3}
\otimes\sigma_{1},\\
\cE_{m+1}&=&\sigma_{2}\otimes\boldsymbol{1}_2\otimes\cdots\otimes\boldsymbol{1}_2,\\
\cE_{m+2}&=&\sigma_{3}\otimes\sigma_{2}\otimes\boldsymbol{1}_2\otimes\cdots\otimes
\boldsymbol{1}_2,\\
\hdotsfor[2]{3}\\
\cE_{2m}&=&\sigma_{3}\otimes\sigma_{3}\otimes\cdots\otimes\sigma_{3}
\otimes\sigma_{2},
\end{array}}\end{equation}
where
\[
\ar \sigma_1=\begin{pmatrix} 0 & 1\\
1 & 0\end{pmatrix},\quad \sigma_2=\begin{pmatrix} 0 & -i\\
i & 0\end{pmatrix},\quad\sigma_3=\begin{pmatrix} i & 0\\
0 & -i\end{pmatrix}
\]
are spinor representations of the units of $\C_2$, $\boldsymbol{1}_2$ is the unit $2\times 2$ matrix.\\[0.3cm]
3) Tensor fields
$(l^\prime,l^{\prime\prime})\oplus(\dot{l}^{\prime\prime},\dot{l}^\prime)$.
The fields $(l^\prime,l^{\prime\prime})$ and
$(\dot{l}^{\prime\prime},\dot{l}^\prime)$ are defined within the
arbitrary spin chains (see Fig.\,1 and Fig.\,2). Universal coverings
of these spin chains are constructed within the representations
$\fC^{l_0+l_1-1,l_0-l_1+1}$ and $\fC^{l_0-l_1+1,l_0+l_1-1}$ of
$\spin_+(1,3)$ in the spinspaces $\dS_{2^{k+r}}$ associated with the
algebra $\C_2\otimes\C_2\otimes\cdots\otimes\C_2\bigotimes
\overset{\ast}{\C}_2\otimes
\overset{\ast}{\C}_2\otimes\cdots\otimes\overset{\ast}{\C}_2$. A
relation between the numbers $l_0$, $l_1$ of the Gel'fand-Naimark
basis (\ref{GNB}) and the numbers $k$ and $r$ of the factors $\C_2$
and $\overset{\ast}{\C}_2$ is given by the following formula:
\[
(l_0,l_1)=\left(\frac{k-r}{2},\frac{k+r}{2}+1\right).
\]
Finally, extended automorphisms groups
$\Ext(\C_{2k}\oplus\overset{\ast}{\C}_{2k})$ and
$\Ext(\C_{2k}\otimes\overset{\ast}{\C}_{2k})$ (correspondingly,
$CPT$ groups) can be derived via the same procedure that described
in detail in our previous work \cite{Var04a}.
\end{proof}

\section{The $CPT$ group of the spin-$1/2$ field}
In accordance with the general Fermi-scheme (Fig.\,1) of the
interlocking representations of $\fG_+$ the field
$(1/2,0)\oplus(0,1/2)$ is defined within the following chain:
\[
\dgARROWLENGTH=2.5em
\begin{diagram}
\node{\left(\frac{1}{2},0\right)}\arrow{e,<>}
\node{\left(0,\frac{1}{2}\right)}
\end{diagram}.
\]
A double covering of the representation associated with the field
$(1/2,0)\oplus(0,1/2)$ is realized in the spinspace
$\dS_2\oplus\dot{\dS}_2$. This spinspace is a space of the
representation $\fC^{1,0}\oplus\fC^{0,-1}$ of $\spin_+(1,3)$.
Further, the algebra $\C_2\oplus\overset{\ast}{\C}_2$ corresponds to
$\fC^{1,0}\oplus\fC^{0,-1}$ and the automorphisms of this algebra
are realized within the representations of $\pin(1,3)$, that is,
they are outer automorphisms with respect to the transformations of
the group $\spin_+(1,3)$. The spinor representations of the
automorphisms, defined on the spinspace $\dS_2\oplus\dot{\dS}_2$,
are constructed via the Brauer-Weyl representation (\ref{BWR}). The
spinbasis of the algebra $\C_2\oplus\overset{\ast}{\C}_2$ is defined
by the following $4\times 4$ matrices:
\[
\cE_1=\sigma_1\otimes\boldsymbol{1}_2=\begin{pmatrix} 0 &
\boldsymbol{1}_2\\
\boldsymbol{1}_2 & 0
\end{pmatrix},\quad
\cE_2=\sigma_3\otimes\sigma_1=\begin{pmatrix} i\sigma_1 & 0\\
0 & -i\sigma_1
\end{pmatrix},
\]
\begin{equation}\label{Basis4}
\cE_3=\sigma_2\otimes\boldsymbol{1}_2=\begin{pmatrix} 0 &
-i\boldsymbol{1}_2\\
i\boldsymbol{1}_2
\end{pmatrix},
\cE_4=\sigma_3\otimes\sigma_2=\begin{pmatrix} i\sigma_2 & 0\\
0 & -i\sigma_2
\end{pmatrix}.
\end{equation}
In accordance with (\ref{SpinW}) we have for the matrix of the
automorphism $\cA\rightarrow\cA^\star$ the following expression:
\[
\sW=\cE_1\cE_2\cE_3\cE_4=\cE_{1234}\sim P.
\]
Further, it is easy to see that among the matrices of the basis
(\ref{Basis4}) there are symmetric and skewsymmetric matrices:
\[
\cE^{\sT}_1=\cE_1,\quad\cE^{\sT}_2=\cE_2,\quad\cE^{\sT}_3=-\cE_3,\quad\cE^{\sT}_4=-\cE_4.
\]
In accordance with $\widetilde{\sA}=\sE\sA^{\sT}\sE^{-1}$ (see
(\ref{SpinE})) we have
\[
\cE_1=\sE\cE_1\sE^{-1},\quad\cE_2=\sE\cE_2\sE^{-1},\quad\cE_3=-\sE\cE_3\sE^{-1},\quad\cE_4=-\sE\cE_4\sE^{-1}.
\]
Hence it follows that $\sE$ commutes with $\cE_1$ and $\cE_2$ and
anticommutes with $\cE_3$ and $\cE_4$, that is, $\sE=\cE_3\cE_4\sim
T$. From the definition $\sC=\sE\sW$ (see (\ref{SpinC})) we find
that the matrix of the antiautomorphism
$\cA\rightarrow\widetilde{\cA^\star}$ has the form
$\sC=\cE_1\cE_2\sim PT$. The basis (\ref{Basis4}) contains both
complex and real matrices:
\[
\cE^\ast_1=\cE_1,\quad\cE^\ast_2=-\cE_2,\quad\cE^\ast_3=-\cE_3,\quad\cE^\ast_4=\cE_4.
\]
Therefore, from $\overline{\sA}=\Pi\sA^\ast\Pi^{-1}$ (see
(\ref{SpinP})) we have
\[
\cE_1=\Pi\cE_1\Pi^{-1},\quad\cE_2=-\Pi\cE_2\Pi^{-1},\quad\cE_3=-\Pi\cE_3\Pi^{-1},\quad\cE_4=\Pi\cE_4\Pi^{-1}.
\]
From the latter relations we obtain $\Pi=\cE_2\cE_3\sim C$. Further,
in accordance with $\sK=\Pi\sW$ (the definition (\ref{SpinK})) for
the matrix of the pseudoautomorphism
$\cA\rightarrow\overline{\cA^\star}$ we have $\sK=\cE_1\cE_4\sim
CP$. Finally, for the pseudoantiautomorphisms
$\cA\rightarrow\overline{\widetilde{\cA}}$ and
$\cA\rightarrow\overline{\widetilde{\cA^\star}}$ from the
definitions $\sS=\Pi\sE$ and $\sF=\Pi\sC$ (see (\ref{SpinS}) and
(\ref{SpinF})) it follows that
$\sS=\cE_2\cE_3\cE_3\cE_4=\cE_2\cE_4\sim CT$ and
$\sF=\cE_2\cE_3\cE_1\cE_2=\cE_1\cE_3\sim CPT$. Thus, we come to the
following automorphism group:
\begin{multline}
\Ext(\C_4)=\{\sI,\sW,\sE,\sC,\Pi,\sK,\sS,\sF\}\simeq\{1,P,T,PT,C,CP,CT,CPT\}\simeq\\
\{\boldsymbol{1}_4,\,\cE_1\cE_2\cE_3\cE_4,\,\cE_3\cE_4,\,\cE_1\cE_2,\,\cE_2\cE_3,\,\cE_1\cE_4,\,\cE_2\cE_4,\,
\cE_1\cE_4\}.\nonumber
\end{multline}
The multiplication table of this group is shown in Tab.\,4. From
this table it follows that $\Ext(\C_4)\simeq D_4$, and for the $CPT$
group we have the following isomorphism: $C^{+,+,+,+,+,-,-}\simeq
D_4\otimes\dZ_2$.
\begin{figure}[ht]
\begin{center}{\renewcommand{\arraystretch}{1.4}
\begin{tabular}{|c||c|c|c|c|c|c|c|c|}\hline
  & $\boldsymbol{1}_4$ & $\cE_{1234}$ & $\cE_{34}$ & $\cE_{12}$ & $\cE_{23}$ &
$\cE_{14}$ & $\cE_{24}$ & $\cE_{13}$\\ \hline\hline
$\boldsymbol{1}_4$ & $\boldsymbol{1}_4$ & $\cE_{1234}$ & $\cE_{34}$
&
$\cE_{12}$ & $\cE_{23}$ & $\cE_{14}$ & $\cE_{24}$ & $\cE_{13}$\\
\hline $\cE_{1234}$ & $\cE_{1234}$ & $\boldsymbol{1}_4$ & $\cE_{12}$
&
$\cE_{34}$ & $\cE_{14}$ & $\cE_{23}$ & $\cE_{13}$ & $\cE_{24}$\\
\hline $\cE_{34}$ & $\cE_{34}$ & $-\cE_{12}$ & $\boldsymbol{1}_4$ &
$\cE_{1234}$
& $-\cE_{24}$ & $-\cE_{13}$ & $-\cE_{23}$ & $-\cE_{14}$\\
\hline $\cE_{12}$ & $\cE_{12}$ & $\cE_{34}$ & $\cE_{1234}$ &
 $\boldsymbol{1}_4$ & $-\cE_{13}$ & $-\cE_{24}$ & $-\cE_{14}$ &
$-\cE_{23}$\\ \hline $\cE_{23}$ & $\cE_{23}$ & $\cE_{14}$ &
$\cE_{24}$ & $\cE_{13}$ & $\boldsymbol{1}_4$ & $\cE_{1234}$ &
$\cE_{34}$ & $\cE_{12}$\\ \hline $\cE_{14}$ & $\cE_{14}$ &
$\cE_{23}$ & $\cE_{13}$ & $\cE_{24}$ & $\cE_{1234}$ &
$\boldsymbol{1}_4$ & $\cE_{12}$ & $\cE_{34}$\\ \hline $\cE_{24}$ &
$\cE_{24}$ & $\cE_{13}$ & $\cE_{23}$
& $\cE_{14}$ & $-\cE_{34}$ & $-\cE_{12}$ & $-\boldsymbol{1}_4$ & $-\cE_{1234}$\\
\hline $\cE_{13}$ & $\cE_{13}$ & $\cE_{24}$ & $\cE_{14}$ &
$\cE_{23}$ & $-\cE_{12}$ & $-\cE_{34}$ & $-\cE_{1234}$ & $-\boldsymbol{1}_4$\\
\hline
\end{tabular}
}
\end{center}
\hspace{0.3cm}
\begin{center}{\small \textbf{Tab.\,4:} The multiplication table of the $CPT/\dZ_2$
group of the field $(1/2,0)\oplus(0,1/2)$.}
\end{center}
\end{figure}

\section{The $CPT$ group of the spin-$1$ field}
In accordance with the general Bose-scheme of the interlocking
representations of $\fG_+$ (see Fig.\,1), the field
$(1,0)\oplus(0,1)$ is defined within the following interlocking
scheme:
\[
\dgARROWLENGTH=2.5em
\begin{diagram}
\node{(1,0)}\arrow{e,<>}\node{\left(\frac{1}{2},\frac{1}{2}\right)}
\arrow{e,<>}\node{(0,1)}
\end{diagram}.
\]
A double covering of the representation, associated with the field
$(1,0)\oplus(0,1)$, is realized in the spinspace
\begin{equation}\label{Spin_1}
\dS_2\otimes\dS_2\bigoplus\dot{\dS}_2\otimes\dot{\dS}_2,
\end{equation}
This spinspace is a space of the representation
$\fC^{2,0}\oplus\fC^{0,-2}$ of the group $\spin_+(1,3)$. The algebra
\begin{equation}\label{Alg_1}
\C_2\otimes\C_2\bigoplus\overset{\ast}{\C}_2\otimes\overset{\ast}{\C}_2.
\end{equation}
is associated with $\fC^{2,0}\oplus\fC^{0,-2}$. The automorphisms of
this algebra are realized within representations of the group
$\pin(1,3)$, that is, they are outer automorphisms with respect
transformations of the group $\spin_+(1,3)$. Spinor representations
of the automorphisms, defined on the spinspace (\ref{Spin_1}), are
constructed via the Brauer-Weyl representation (\ref{BWR}). A
spinbasis of the algebra (\ref{Alg_1}) is defined by the following
$8\times 8$ matrices:
\[
\cE_1=\sigma_1\otimes\boldsymbol{1}_2\otimes\boldsymbol{1}_2=\begin{bmatrix}
0 & 0 & \boldsymbol{1}_2 & 0\\
0 & 0 & 0 & \boldsymbol{1}_2\\
\boldsymbol{1}_2 & 0 & 0 & 0\\
0 & \boldsymbol{1}_2 & 0 & 0
\end{bmatrix},
\]
\[
\cE_2=\sigma_3\otimes\sigma_1\otimes\boldsymbol{1}_2=\begin{bmatrix}
0 & i\boldsymbol{1}_1 & 0 & 0\\
i\boldsymbol{1}_2 & 0 & 0 & 0\\
0 & 0 & 0 & -i\boldsymbol{1}_2\\
0 & 0 & -i\boldsymbol{1}_2 & 0
\end{bmatrix},
\]
\[
\cE_3=\sigma_3\otimes\sigma_3\otimes\sigma_1=\begin{bmatrix}
-\sigma_1 & 0 & 0 & 0\\
0 & \sigma_1 & 0 & 0\\
0 & 0 & \sigma_1 & 0\\
0 & 0 & 0 & -\sigma_1
\end{bmatrix},
\]
\[
\cE_4=\sigma_2\otimes\boldsymbol{1}_2\otimes\boldsymbol{1}_2=\begin{bmatrix}
0 & 0 & -i\boldsymbol{1}_2 & 0\\
0 & 0 & 0 & -i\boldsymbol{1}_2\\
i\boldsymbol{1}_2 & 0 & 0 & 0\\
0 & i\boldsymbol{1}_2 & 0 & 0
\end{bmatrix},
\]
\[
\cE_5=\sigma_3\otimes\sigma_2\otimes\boldsymbol{1}_2=\begin{bmatrix}
0 & \boldsymbol{1}_2 & 0 & 0\\
-\boldsymbol{1}_2 & 0 & 0 & 0\\
0 & 0 & 0 & -\boldsymbol{1}_2\\
0 & 0 & \boldsymbol{1}_2 & 0
\end{bmatrix},
\]
\[
\cE_6=\sigma_3\otimes\sigma_3\otimes\sigma_2=\begin{bmatrix}
-\sigma_2 & 0 & 0 & 0\\
0 & \sigma_2 & 0 & 0\\
0 & 0 & \sigma_2 & 0\\
0 & 0 & 0 & -\sigma_2
\end{bmatrix}.
\]
Using these matrices, we construct $CPT$ group for the field
$(1,0)\oplus(0,1)$. At first, the matrix of the automorphism
$\cA\rightarrow\cA^\star$ has the form
\[
\sW=\cE_1\cE_2\cE_3\cE_4\cE_5\cE_6=\cE_{123456}\sim P.
\]
Further, since
\[
\cE^{\sT}_1=\cE_1,\quad\cE^{\sT}_2=\cE_2,\quad\cE^{\sT}_3=\cE_3,\quad\cE^{\sT}_4=-\cE_4,
\quad\cE^{\sT}_5=-\cE_5,\quad\cE^{\sT}_6=-\cE_6,
\]
then in accordance with $\widetilde{\sA}=\sE\sA^{\sT}\sE^{-1}$ we
have
\[
\cE_1=\sE\cE_1\sE^{-1},\quad\cE_2=\sE\cE_2\sE^{-1},\quad\cE_3=\sE\cE_3\sE^{-1},\quad
\cE_4=-\sE\cE_4\sE^{-1},
\]
\[
\cE_5=-\sE\cE_5\sE^{-1},\quad\cE_6=-\sE\cE_6\sE^{-1}.
\]
Hence it follows that $\sE$ commutes with $\cE_1$, $\cE_2$, $\cE_3$
and anticommutes with $\cE_4$, $\cE_5$, $\cE_6$, that is,
$\sE=\cE_{456}\sim T$. From the definition $\sC=\sE\sW$ we find that
a matrix of the antiautomorphism
$\cA\rightarrow\widetilde{\cA^\star}$ has the form
$\sC=\cE_{123}\sim PT$. The basis
$\{\cE_1,\cE_2,\cE_3,\cE_4,\cE_5,\cE_6\}$ contains both complex and
real matrices:
\[
\cE^\ast_1=\cE_1,\quad\cE^\ast_2=-\cE_2,\quad\cE^\ast_3=\cE_3,\quad\cE^\ast_4=-\cE_4,\quad
\cE^\ast_5=\cE_5,\quad\cE^\ast_6=-\cE_6.
\]
Therefore, from $\overline{\sA}=\Pi\sA^\ast\Pi^{-1}$ we have
\[
\cE_1=\Pi\cE_1\Pi^{-1},\quad\cE_2=-\Pi\cE_2\Pi^{-1},\quad\cE_3=\Pi\cE_3\Pi^{-1},\quad
\cE_4=-\Pi\cE_4\Pi^{-1},
\]
\[
\cE_5=\Pi\cE_5\Pi^{-1},\quad\cE_6=-\Pi\cE_6\Pi^{-1}.
\]
From the latter relations we obtain $\Pi=\cE_{246}\sim C$. Further,
in accordance with $\sK=\Pi\sW$ for the matrix of the
pseudoautomorphism $\cA\rightarrow\overline{\cA^\star}$ we have
$\sK=\cE_{135}\sim CP$. Finally, for the pseudoantiautomorphisms
$\cA\rightarrow\overline{\widetilde{\cA}}$,
$\cA\rightarrow\overline{\widetilde{\cA^\star}}$ from the
definitions $\sS=\Pi\sE$, $\sF=\Pi\sC$ it follows that
$\sS=\cE_{25}\sim CT$, $\sF=\cE_{1346}\sim CPT$. Thus, we come to
the following automorphism group:
\begin{multline}
\Ext(\C_6)\simeq\{\sI,\sW,\sE,\sC,\Pi,\sK,\sS,\sF\}\simeq\{1,P,T,PT,C,CP,CT,CPT\}\simeq\\
\{\boldsymbol{1}_8,\cE_{123456},\cE_{456},\cE_{123},\cE_{246},\cE_{135},\cE_{25},\cE_{1346}\}.
\nonumber
\end{multline}
The multiplication table of this group is given in Tab.\,5. From
this table it follows that $\Ext(\C_6)\simeq D_4$, and for the $CPT$
group we have the following isomorphism: $C^{-,+,+,+,+,-,+}\simeq
D_4\otimes\dZ_2$.
\begin{figure}[ht]
\begin{center}{\renewcommand{\arraystretch}{1.4}
\begin{tabular}{|c||c|c|c|c|c|c|c|c|}\hline
  & $\boldsymbol{1}_8$ & $\cE_{123456}$ & $\cE_{456}$ & $\cE_{123}$ & $\cE_{246}$ &
$\cE_{135}$ & $\cE_{25}$ & $\cE_{1346}$\\ \hline\hline
$\boldsymbol{1}_8$ & $\boldsymbol{1}_8$ & $\cE_{123456}$ &
$\cE_{456}$ & $\cE_{123}$ & $\cE_{246}$ & $\cE_{135}$ & $\cE_{25}$ &
$\cE_{1346}$\\ \hline $\cE_{123456}$ & $\cE_{123456}$ &
$-\boldsymbol{1}_8$ & $\cE_{123}$ & $-\cE_{456}$ & $-\cE_{135}$ &
$\cE_{246}$ & $-\cE_{1346}$ & $\cE_{25}$\\ \hline $\cE_{456}$ &
$\cE_{456}$ & $-\cE_{123}$ & $\boldsymbol{1}_8$ & $-\cE_{123456}$
& $-\cE_{25}$ & $\cE_{1346}$ & $-\cE_{246}$ & $\cE_{135}$\\
\hline $\cE_{123}$ & $\cE_{123}$ & $\cE_{456}$ & $\cE_{123456}$ &
 $\boldsymbol{1}_8$ & $\cE_{1346}$ & $\cE_{25}$ & $\cE_{135}$ &
$\cE_{246}$\\ \hline $\cE_{246}$ & $\cE_{246}$ & $\cE_{135}$ &
$\cE_{25}$ & $\cE_{1346}$ & $\boldsymbol{1}_8$ & $\cE_{123456}$ &
$\cE_{456}$ & $\cE_{123}$\\ \hline $\cE_{135}$ & $\cE_{135}$ &
$-\cE_{246}$ & $\cE_{1346}$ & $-\cE_{25}$ & $-\cE_{123456}$ &
$\boldsymbol{1}_8$ & $-\cE_{123}$ & $\cE_{456}$\\ \hline $\cE_{25}$
& $\cE_{25}$ & $-\cE_{1346}$ & $\cE_{246}$
& $-\cE_{135}$ & $-\cE_{456}$ & $\cE_{123}$ & $-\boldsymbol{1}_8$ & $\cE_{123456}$\\
\hline $\cE_{1346}$ & $\cE_{1346}$ & $\cE_{25}$ & $\cE_{135}$ &
$\cE_{246}$ & $\cE_{123}$ & $\cE_{456}$ & $\cE_{123456}$ & $\boldsymbol{1}_8$\\
\hline
\end{tabular}
}
\end{center}
\hspace{0.3cm}
\begin{center}{\small \textbf{Tab.\,5:} The multiplication table of the $CPT/\dZ_2$
group of the field $(1,0)\oplus(0,1)$.}
\end{center}
\end{figure}

\section{The $CPT$ group of the spin-$3/2$ field}
In accordance with the general Fermi-scheme of the interlocking
representations of $\fG_+$ (see Fig.\,2), the field
$(3/2,0)\oplus(0,3/2)$ is defined within the following interlocking
scheme:
\[
\dgARROWLENGTH=2.5em
\begin{diagram}
\node{\left(\frac{3}{2},0\right)}\arrow{e,<>}\node{\left(1,\frac{1}{2}\right)}
\arrow{e,<>}\node{\left(\frac{1}{2},1\right)}\arrow{e,<>}\node{\left(0,\frac{3}{2}\right)}
\end{diagram}.
\]
A double covering of the representation, associated with the field
$(3/2,0)\oplus(0,3/2)$, is realized in the spinspace
\begin{equation}\label{Spin_3}
\dS_2\otimes\dS_2\otimes\dS_2\bigoplus\dot{\dS}_2\otimes\dot{\dS}_2\otimes\dot{\dS}_2,
\end{equation}
This spinspace is a space of the representation
$\fC^{3,0}\oplus\fC^{0,-3}$ of the group $\spin_+(1,3)$. The algebra
\begin{equation}\label{Alg_3}
\C_2\otimes\C_2\otimes\C_2\bigoplus\overset{\ast}{\C}_2\otimes\overset{\ast}{\C}_2\otimes\overset{\ast}{\C}_2
\end{equation}
is associated with the representation $\fC^{3,0}\oplus\fC^{0,-3}$.
Spinor representations of the automorphisms, defined on the
spinspace (\ref{Spin_3}), are constructed via the Brauer-Weyl
representation (\ref{BWR}). A spinbasis of the algebra (\ref{Alg_3})
is defined by the following $16\times 16$ matrices:
\[
\cE_1=\sigma_1\otimes\boldsymbol{1}_2\otimes\boldsymbol{1}_2\otimes\boldsymbol{1}_2=
\begin{bmatrix}
0 & 0 & 0 & 0 & \boldsymbol{1}_2 & 0 & 0 & 0\\
0 & 0 & 0 & 0 & 0 & \boldsymbol{1}_2 & 0 & 0\\
0 & 0 & 0 & 0 & 0 & 0 & \boldsymbol{1}_2 & 0\\
0 & 0 & 0 & 0 & 0 & 0 & 0 & \boldsymbol{1}_2\\
\boldsymbol{1}_2  & 0 & 0 & 0 & 0 & 0 & 0 & 0\\
0 & \boldsymbol{1}_2  & 0 & 0 & 0 & 0 & 0 & 0\\
0 & 0 & \boldsymbol{1}_2  & 0 & 0 & 0 & 0 & 0\\
0 & 0 & 0 & \boldsymbol{1}_2  & 0 & 0 & 0 & 0
\end{bmatrix},
\]
\[
\cE_2=\sigma_3\otimes\sigma_1\otimes\boldsymbol{1}_2\otimes\boldsymbol{1}_2=
\begin{bmatrix}
0 & 0 & i\boldsymbol{1}_2 & 0 & 0 & 0 & 0 & 0\\
0 & 0 & 0 & i\boldsymbol{1}_2 & 0 & 0 & 0 & 0\\
i\boldsymbol{1}_2 & 0 & 0 & 0 & 0 & 0 & 0 & 0\\
0 & i\boldsymbol{1}_2 & 0 & 0 & 0 & 0 & 0 & 0\\
0 & 0 & 0 & 0 & 0 & 0 &-i\boldsymbol{1}_2 & 0\\
0 & 0 & 0 & 0 & 0 & 0 & 0 &-i\boldsymbol{1}_2\\
0 & 0 & 0 & 0 &-i\boldsymbol{1}_2 & 0 & 0 & 0\\
0 & 0 & 0 & 0 & 0 &-i\boldsymbol{1}_2 & 0 & 0
\end{bmatrix},
\]
\[
\cE_3=\sigma_3\otimes\sigma_3\otimes\sigma_1\otimes\boldsymbol{1}_2=
\begin{bmatrix}
0 &-\boldsymbol{1}_2 & 0 & 0 & 0 & 0 & 0 & 0\\
-\boldsymbol{1}_2 & 0 & 0 & 0 & 0 & 0 & 0 & 0\\
0 & 0 & 0 & \boldsymbol{1}_2 & 0 & 0 & 0 & 0\\
0 & 0 & \boldsymbol{1}_2 & 0 & 0 & 0 & 0 & 0\\
0 & 0 & 0 & 0 & 0 & \boldsymbol{1}_2 & 0 & 0\\
0 & 0 & 0 & 0 & \boldsymbol{1}_2 & 0 & 0 & 0\\
0 & 0 & 0 & 0 & 0 & 0 & 0 &-\boldsymbol{1}_2\\
0 & 0 & 0 & 0 & 0 & 0 &-\boldsymbol{1}_2 & 0
\end{bmatrix},
\]
\[
\cE_4=\sigma_3\otimes\sigma_3\otimes\sigma_3\otimes\sigma_1=
\begin{bmatrix}
-i\sigma_1 & 0 & 0 & 0 & 0 & 0 & 0 & 0\\
0 & i\sigma_1 & 0 & 0 & 0 & 0 & 0 & 0\\
0 & 0 & i\sigma_1 & 0 & 0 & 0 & 0 & 0\\
0 & 0 & 0 &-i\sigma_1 & 0 & 0 & 0 & 0\\
0 & 0 & 0 & 0 & i\sigma_1 & 0 & 0 & 0\\
0 & 0 & 0 & 0 & 0 &-i\sigma_1 & 0 & 0\\
0 & 0 & 0 & 0 & 0 & 0 &-i\sigma_1 & 0\\
0 & 0 & 0 & 0 & 0 & 0 & 0 & i\sigma_1
\end{bmatrix},
\]
\[
\cE_5=\sigma_2\otimes\boldsymbol{1}_2\otimes\boldsymbol{1}_2\otimes\boldsymbol{1}_2=
\begin{bmatrix}
0 & 0 & 0 & 0 &-i\boldsymbol{1}_2 & 0 & 0 & 0\\
0 & 0 & 0 & 0 & 0 &-i\boldsymbol{1}_2 & 0 & 0\\
0 & 0 & 0 & 0 & 0 & 0 &-i\boldsymbol{1}_2 & 0\\
0 & 0 & 0 & 0 & 0 & 0 & 0 &-i\boldsymbol{1}_2\\
i\boldsymbol{1}_2 & 0 & 0 & 0 & 0 & 0 & 0 & 0\\
0 & i\boldsymbol{1}_2 & 0 & 0 & 0 & 0 & 0 & 0\\
0 & 0 & i\boldsymbol{1}_2 & 0 & 0 & 0 & 0 & 0\\
0 & 0 & 0 & i\boldsymbol{1}_2 & 0 & 0 & 0 & 0
\end{bmatrix},
\]
\[
\cE_6=\sigma_3\otimes\sigma_2\otimes\boldsymbol{1}_2\otimes\boldsymbol{1}_2=
\begin{bmatrix}
0 & 0 & i\boldsymbol{1}_2 & 0 & 0 & 0 & 0 & 0\\
0 & 0 & 0 & i\boldsymbol{1}_2 & 0 & 0 & 0 & 0\\
-i\boldsymbol{1}_2& 0 & 0 & 0 & 0 & 0 & 0 & 0\\
0 &-i\boldsymbol{1}_2 & 0 & 0 & 0 & 0 & 0 & 0\\
0 & 0 & 0 & 0 & 0 & 0 &-i\boldsymbol{1}_2 & 0\\
0 & 0 & 0 & 0 & 0 & 0 & 0 &-i\boldsymbol{1}_2\\
0 & 0 & 0 & 0 & i\boldsymbol{1}_2 & 0 & 0 & 0\\
0 & 0 & 0 & 0 & 0 & i\boldsymbol{1}_2 & 0 & 0
\end{bmatrix},
\]
\[
\cE_7=\sigma_3\otimes\sigma_3\otimes\sigma_2\otimes\boldsymbol{1}_2=
\begin{bmatrix}
0 & i\boldsymbol{1}_2 & 0 & 0 & 0 & 0 & 0 & 0\\
-i\boldsymbol{1}_2& 0 & 0 & 0 & 0 & 0 & 0 & 0\\
0 & 0 & 0 &-i\boldsymbol{1}_2 & 0 & 0 & 0 & 0\\
0 & 0 & i\boldsymbol{1}_2 & 0 & 0 & 0 & 0 & 0\\
0 & 0 & 0 & 0 & 0 &-i\boldsymbol{1}_2 & 0 & 0\\
0 & 0 & 0 & 0 & i\boldsymbol{1}_2 & 0 & 0 & 0\\
0 & 0 & 0 & 0 & 0 & 0 & 0 & i\boldsymbol{1}_2\\
0 & 0 & 0 & 0 & 0 & 0 &-i\boldsymbol{1}_2 & 0
\end{bmatrix},
\]
\[
\cE_8=\sigma_3\otimes\sigma_3\otimes\sigma_3\otimes\sigma_2=
\begin{bmatrix}
-i\sigma_2& 0 & 0 & 0 & 0 & 0 & 0 & 0\\
0 & i\sigma_2 & 0 & 0 & 0 & 0 & 0 & 0\\
0 & 0 & i\sigma_2 & 0 & 0 & 0 & 0 & 0\\
0 & 0 & 0 &-i\sigma_2 & 0 & 0 & 0 & 0\\
0 & 0 & 0 & 0 & i\sigma_2 & 0 & 0 & 0\\
0 & 0 & 0 & 0 & 0 &-i\sigma_2 & 0 & 0\\
0 & 0 & 0 & 0 & 0 & 0 &-i\sigma_2 & 0\\
0 & 0 & 0 & 0 & 0 & 0 & 0 & i\sigma_2
\end{bmatrix}.
\]
Using this spinbasis, we construct $CPT$ group for the field
$(3/2,0)\oplus(0,3/2)$. At first, the matrix of the automorphism
$\cA\rightarrow\cA^\star$ has the form
\[
\sW=\cE_1\cE_2\cE_3\cE_4\cE_5\cE_6\cE_7\cE_8=\cE_{12345678}\sim P.
\]
Further, since
\[
\cE^{\sT}_1=\cE_1,\quad\cE^{\sT}_2=\cE_2,\quad\cE^{\sT}_3=\cE_3,\quad\cE^{\sT}_4=\cE_4,
\quad\cE^{\sT}_5=-\cE_5,\quad\cE^{\sT}_6=-\cE_6,\quad\cE^{\sT}_7=-\cE_7,\quad
\cE^{\sT}_8=-\cE_8,
\]
then in accordance with $\widetilde{\sA}=\sE\sA^{\sT}\sE^{-1}$ we
have
\[
\cE_1=\sE\cE_1\sE^{-1},\quad\cE_2=\sE\cE_2\sE^{-1},\quad\cE_3=\sE\cE_3\sE^{-1},\quad
\cE_4=\sE\cE_4\sE^{-1},
\]
\[
\cE_5=-\sE\cE_5\sE^{-1},\quad\cE_6=-\sE\cE_6\sE^{-1},\quad
\cE_7=-\sE\cE_7\sE^{-1},\quad\cE_8=-\sE\cE_8\sE^{-1}.
\]
Hence it follows that $\sE$ commutes with $\cE_1$, $\cE_2$, $\cE_3$,
$\cE_4$ and anticommutes with $\cE_5$, $\cE_6$, $\cE_7$, $\cE_8$,
that is, $\sE=\cE_{5678}\sim T$. From the definition $\sC=\sE\sW$ we
find that a matrix of the antiautomorphism
$\cA\rightarrow\widetilde{\cA^\star}$ has the form
$\sC=\cE_{1234}\sim PT$. The basis
$\{\cE_1,\cE_2,\cE_3,\cE_4,\cE_5,\cE_6,\cE_7,\cE_8\}$ contains both
complex and real matrices:
\[
\cE^\ast_1=\cE_1,\quad\cE^\ast_2=-\cE_2,\quad\cE^\ast_3=\cE_3,\quad\cE^\ast_4=-\cE_4,\quad
\cE^\ast_5=-\cE_5,\quad\cE^\ast_6=-\cE_6,\quad\cE^\ast_7=-\cE_7,\quad\cE^\ast_8=\cE_8.
\]
Therefore, from $\overline{\sA}=\Pi\sA^\ast\Pi^{-1}$ we have
\[
\cE_1=\Pi\cE_1\Pi^{-1},\quad\cE_2=-\Pi\cE_2\Pi^{-1},\quad\cE_3=\Pi\cE_3\Pi^{-1},\quad
\cE_4=-\Pi\cE_4\Pi^{-1},
\]
\[
\cE_5=-\Pi\cE_5\Pi^{-1},\quad\cE_6=-\Pi\cE_6\Pi^{-1},\quad\cE_7=-\Pi\cE_7\Pi^{-1},\quad
\cE_8=\Pi\cE_8\Pi^{-1}.
\]
From the latter relations we obtain $\Pi=\cE_{24567}\sim C$.
Further, in accordance with $\sK=\Pi\sW$ for the matrix of the
pseudoautomorphism $\cA\rightarrow\overline{\cA^\star}$ we have
$\sK=\cE_{138}\sim CP$. Finally, for the pseudoantiautomorphisms
$\cA\rightarrow\overline{\widetilde{\cA}}$ and
$\cA\rightarrow\overline{\widetilde{\cA^\star}}$ from the
definitions $\sS=\Pi\sE$ and $\sF=\Pi\sC$ it follows that
$\sS=\cE_{248}\sim CT$, $\sF=\cE_{13567}\sim CPT$. Thus, we come to
the following automorphism group:
\begin{multline}
\Ext(\C_8)\simeq\{\sI,\sW,\sE,\sC,\Pi,\sK,\sS,\sF\}\simeq\{1,P,T,PT,C,CP,CT,CPT\}\simeq\\
\{\boldsymbol{1}_{16},\cE_{12345678},\cE_{5678},\cE_{1234},\cE_{24567},\cE_{138},\cE_{246},\cE_{13567}\}.
\nonumber
\end{multline}
The multiplication table of this group is given in Tab.\,6. From
this table it follows that $\Ext(\C_8)\simeq D_4$, and for the $CPT$
group we have the following isomorphism: $C^{-,-,+,+,+,+,+}\simeq
D_4\otimes\dZ_2$.
\begin{figure}[ht]
\begin{center}{\renewcommand{\arraystretch}{1.4}
\begin{tabular}{|c||c|c|c|c|c|c|c|c|}\hline
  & $\boldsymbol{1}_{16}$ & $\sW$ & $\cE_{5678}$ & $\cE_{1234}$ & $\cE_{24567}$ &
$\cE_{138}$ & $\cE_{248}$ & $\cE_{13567}$\\ \hline\hline
$\boldsymbol{1}_{16}$ & $\boldsymbol{1}_{16}$ & $\sW$ & $\cE_{5678}$
& $\cE_{1234}$ & $\cE_{24567}$ & $\cE_{138}$ & $\cE_{248}$ &
$\cE_{13567}$\\ \hline $\sW$ & $\sW$ & $-\boldsymbol{1}_{16}$ &
$-\cE_{1234}$ & $\cE_{5678}$ &
$\cE_{138}$ & $-\cE_{24567}$ & $-\cE_{13567}$ & $-\cE_{248}$\\
\hline $\cE_{5678}$ & $\cE_{5678}$ & $-\cE_{1234}$ &
$-\boldsymbol{1}_{16}$ & $\sW$
& $\cE_{248}$ & $-\cE_{13567}$ & $-\cE_{24567}$ & $\cE_{138}$\\
\hline $\cE_{1234}$ & $\cE_{1234}$ & $\cE_{5678}$ & $\sW$ &
 $\boldsymbol{1}_{16}$ & $\cE_{13567}$ & $\cE_{248}$ & $\cE_{138}$ &
$\cE_{24567}$\\ \hline $\cE_{24567}$ & $\cE_{24567}$ & $-\cE_{138}$
& $-\cE_{248}$ & $\cE_{13567}$ & $\boldsymbol{1}_{16}$ & $-\sW$ &
$-\cE_{5678}$ & $\cE_{1234}$\\ \hline $\cE_{138}$ & $\cE_{138}$ &
$\cE_{24567}$ & $\cE_{13567}$ & $\cE_{248}$ & $\sW$ &
$\boldsymbol{1}_{16}$ & $\cE_{1234}$ & $\cE_{5678}$\\ \hline
$\cE_{248}$ & $\cE_{248}$ & $\cE_{13567}$ & $\cE_{24567}$
& $\cE_{138}$ & $\cE_{5678}$ & $\cE_{1234}$ & $\boldsymbol{1}_{16}$ & $\sW$\\
\hline $\cE_{13567}$ & $\cE_{13567}$ & $-\cE_{248}$ & $-\cE_{138}$ &
$\cE_{24567}$ & $\cE_{1234}$ & $-\cE_{5678}$ & $-\sW$ & $\boldsymbol{1}_{16}$\\
\hline
\end{tabular}
}
\end{center}
\hspace{0.3cm}
\begin{center}{\small \textbf{Tab.\,6:} The multiplication table of the $CPT/\dZ_2$
group of the field $(3/2,0)\oplus(0,3/2)$.}
\end{center}
\end{figure}

\section{$CPT$ groups of the tensor fields}
As it is shown in the section 3 double coverings of the
representations associated with the tensor fields are constructed
within the product
$\C_2\otimes\C_2\otimes\cdots\otimes\C_2\bigotimes
\overset{\ast}{\C}_2\otimes\overset{\ast}{\C}_2\otimes\cdots\otimes
\overset{\ast}{\C}_2$, where we have $k$ algebras $\C_2$ and $r$
algebras $\overset{\ast}{\C}_2$. A relation between the number $l$
(a weight of the representation in the Van der Waerden basis
(\ref{Waerden})) and the numbers $k$ and $r$ is given by the formula
\begin{equation}\label{LKR}
l=\frac{k-r}{2}.
\end{equation}
It is easy to see that a central row in the scheme shown on the
Fig.\,1,
\begin{equation}\label{BRow}
\dgARROWLENGTH=0.5em \dgHORIZPAD=1.7em \dgVERTPAD=2.2ex
\begin{diagram}
\node{(0,0)}\arrow{e,-}
\node{\left(\frac{1}{2},\frac{1}{2}\right)}\arrow{e,-}
\node{(1,1)}\arrow{e,-}\node{\cdots}\arrow{e,-}
\node{\left(\frac{s}{2},\frac{s}{2}\right)}\arrow{e,-} \node{\cdots}
\end{diagram}
\end{equation}
in virtue of (\ref{LKR}) is equivalent to the following row:
\[
\dgARROWLENGTH=0.5em \dgHORIZPAD=1.7em \dgVERTPAD=2.2ex
\begin{diagram}
\node{[0,0]}\arrow{e,-} \node{\left[0,0\right]}\arrow{e,-}
\node{[0,0]}\arrow{e,-}\node{\cdots}\arrow{e,-}
\node{\left[0,0\right]}\arrow{e,-} \node{\cdots}
\end{diagram}
\]
Analogously, the row shown on the Fig.\,2,
\begin{equation}\label{FRow}
\dgARROWLENGTH=0.5em \dgHORIZPAD=1.7em \dgVERTPAD=2.2ex
\begin{diagram}
\node{\left(\frac{1}{2},0\right)}\arrow{e,-}
\node{\left(1,\frac{1}{2}\right)}\arrow{e,-}
\node{\cdots}\arrow{e,-}
\node{\left(\frac{2s+1}{4},\frac{2s-1}{4}\right)}\arrow{e,-}
\node{\cdots}
\end{diagram}
\end{equation}
is equivalent to
\[
\dgARROWLENGTH=0.5em \dgHORIZPAD=1.7em \dgVERTPAD=2.2ex
\begin{diagram}
\node{\left[\frac{1}{2},0\right]}\arrow{e,-}
\node{\left[\frac{1}{2},0\right]}\arrow{e,-}
\node{\cdots}\arrow{e,-}
\node{\left[\frac{1}{2},0\right]}\arrow{e,-} \node{\cdots}
\end{diagram}
\]
Therefore, all the representations of $\spin_+(1,3)$ can be divided
on the equivalent rows which we show on the Fig.\,3 and Fig.\,4.
\begin{figure}[ht]
\[
\dgARROWLENGTH=0.5em \dgHORIZPAD=1.7em \dgVERTPAD=2.2ex
\begin{diagram}
\node[5]{[s,0]}\arrow{e,-}\arrow{s,-}\node{\cdots}\\
\node[5]{\vdots}\arrow{s,-}\\
\node[3]{[2,0]}\arrow{e,-}\arrow{s,-}\node{\cdots}\arrow{e,-}
\node{\left[2,0\right]}\arrow{s,-}\arrow{e,-}
\node{\cdots}\\
\node[2]{[1,0]}\arrow{s,-}\arrow{e,-}
\node{\left[1,0\right]}\arrow{s,-}\arrow{e,-}
\node{\cdots}\arrow{e,-}
\node{\left[1,0\right]}\arrow{s,-}\arrow{e,-}
\node{\cdots}\\
\node{[0,0]}\arrow{e,-}
\node{\left[0,0\right]}\arrow{s,-}\arrow{e,-}
\node{[0,0]}\arrow{s,-}\arrow{e,-}\node{\cdots}\arrow{e,-}
\node{\left[0,0\right]}\arrow{s,-}\arrow{e,-}
\node{\cdots}\\
\node[2]{[0,1]}\arrow{e,-}
\node{\left[0,1\right]}\arrow{s,-}\arrow{e,-}
\node{\cdots}\arrow{e,-}
\node{\left[0,1\right]}\arrow{s,-}\arrow{e,-}
\node{\cdots}\\
\node[3]{[0,2]}\arrow{e,-}\node{\cdots}\arrow{e,-}
\node{\left[0,2\right]}\arrow{s,-}\arrow{e,-}
\node{\cdots}\\
\node[5]{\vdots}\arrow{s,-}\\
\node[5]{[0,s]}\arrow{e,-}\node{\cdots}
\end{diagram}
\]
\begin{center}{\small {\bf Fig.\,3:} Integer spin representations of the group $\spin_+(1,3)$.}\end{center}
\end{figure}
\begin{figure}[ht]
\[
\dgARROWLENGTH=0.5em
\dgHORIZPAD=1.7em %1.5em
\dgVERTPAD=2.2ex %2ex
\begin{diagram}
\node[4]{[s,0]}\arrow{s,-}\arrow{e,-}\node{\cdots}\\
\node[4]{\vdots}\arrow{s,-}\\
\node[2]{\left[\frac{3}{2},0\right]}\arrow{s,-}\arrow{e,-}
\node{\cdots}\arrow{e,-}
\node{\left[\frac{3}{2},0\right]}\arrow{s,-}\arrow{e,-}
\node{\cdots}\\
\node{\left[\frac{1}{2},0\right]}\arrow{s,-}\arrow{e,-}
\node{\left[\frac{1}{2},0\right]}\arrow{s,-}\arrow{e,-}
\node{\cdots}\arrow{e,-}
\node{\left[\frac{1}{2},0\right]}\arrow{s,-}\arrow{e,-}
\node{\cdots}\\
\node{\left[0,\frac{1}{2}\right]}\arrow{e,-}
\node{\left[0,\frac{1}{2}\right]}\arrow{s,-}\arrow{e,-}
\node{\cdots}\arrow{e,-}
\node{\left[0,\frac{1}{2}\right]}\arrow{s,-}\arrow{e,-}
\node{\cdots}\\
\node[2]{\left[0,\frac{3}{2}\right]}\arrow{e,-}
\node{\cdots}\arrow{e,-}
\node{\left[0,\frac{3}{2}\right]}\arrow{s,-}\arrow{e,-}
\node{\cdots}\\
\node[4]{\vdots}\arrow{s,-}\\
\node[4]{[0,s]}\arrow{e,-}\node{\cdots}
\end{diagram}
\]
\begin{center}{\small {\bf Fig.\,4:} Half-integer spin representations of the group $\spin_+(1,3)$.}\end{center}
\end{figure}
On the other hand, the row (\ref{BRow}) corresponds to the following
chain of the algebras:
\begin{multline}
\boldsymbol{1}\;\longrightarrow\;\C_2\otimes\overset{\ast}{\C}_2\;\longrightarrow\;
\C_2\otimes\C_2\bigotimes\overset{\ast}{\C}_2\otimes\overset{\ast}{\C}_2\;\longrightarrow\;\ldots\;\longrightarrow\\
\longrightarrow\;\underbrace{\C_2\otimes\C_2\otimes\cdots\otimes\C_2}_{s\;\text{times}}\bigotimes
\underbrace{\overset{\ast}{\C}_2\otimes\overset{\ast}{\C}_2
\otimes\cdots\otimes\overset{\ast}{\C}_2}_{s\;\text{times}}\;\longrightarrow\;\ldots
\nonumber
\end{multline}
In its turn, the row (\ref{FRow}) corresponds to the chain
\[
\C_2\;\longrightarrow\;\C_2\otimes\C_2\bigotimes\overset{\ast}{\C}_2\;\longrightarrow\;\ldots\;\longrightarrow
\;\underbrace{\C_2\otimes\C_2\otimes\cdots\otimes\C_2}_{(2s+1)/2\;\text{times}}\bigotimes
\underbrace{\overset{\ast}{\C}_2\otimes\overset{\ast}{\C}_2
\otimes\cdots\otimes\overset{\ast}{\C}_2}_{(2s-1)/2\;\text{times}}\;\longrightarrow\;\ldots
\]
Moreover, these chains induces the following chains of the
spinspaces:
\[
\dS_0\;\longrightarrow\;\dS_4\;\longrightarrow\;\dS_{16}\;\longrightarrow\;\ldots\longrightarrow\;
\dS_{2^{2s}}\;\longrightarrow\;\ldots
\]
and
\[
\dS_2\;\longrightarrow\;\dS_8\;\longrightarrow\;\ldots\;\longrightarrow\;\dS_{2^{2s}}\;\longrightarrow\;\ldots
\]
Thus, the row (\ref{BRow}) (or (\ref{FRow})) induces a sequence of
the fields of the spin 0 (or $1/2$) realized in the spinspaces of
different dimensions. In general case presented on the Fig.\,3 and
Fig.\,4 we have sequences of the fields of the same spin realized in
the different representation spaces of $\spin_+(1,3)$. One can say
that this situation corresponds to particles of the same spin with
different masses, like proton $\longrightarrow$ electron
$\longrightarrow\;\ldots$ (spin $1/2$). With the aim to give more
detailed explanation for this statement let us consider a
Gel'fand-Yaglom mass spectrum formula \cite{GMS}:
\begin{equation}\label{MSF}
\mu^{(l)}=\frac{\kappa}{l+\frac{1}{2}}=\frac{2\kappa}{2l+1},
\end{equation}
where the mass $\mu^{(l)}$ corresponds the spin $l$, $\kappa$ is a
constant. It is easy to see that the denominator $2l+1$ in
(\ref{MSF}) is equal to a dimensionality of the representation space
$\Sym_{(k,0)}$ corresponding to the field
$\boldsymbol{\psi}(\balpha)$ of type $(l,0)$ (or $(0,\dot{l})$ and
$\Sym_{(0,r)}$). For the tensor fields $\boldsymbol{\psi}(\balpha)$
of type $(l\dot{l})$ we have
\begin{equation}\label{MSTF}
\mu^{(s)}=\frac{\kappa}{(k+1)(r+1)},
\end{equation}
where $s=|k-r|/2$ is a spin of the field
$\boldsymbol{\psi}(\balpha)$. In this case, the denominator in
(\ref{MSTF}) is equal to a dimensionality of the representation
space $\Sym_{(k,r)}$ corresponding to the tensor field. Mass
spectrum formulas (\ref{MSF}) and (\ref{MSTF}) give a relationship
between dimensions of the representation spaces of $\spin_+(1,3)$
and particle masses. From the formula (\ref{MSTF}) it follows
directly that on the parallel rows presented on the Fig.\,3 and
Fig.\,4 we have particles of the same spin with different masses.
When $l\to\infty$ (or $(k+1)(r+1)\to\infty$) we come to particles
with zero mass (like a photon). In this case, finite-dimensional
representation spaces $\Sym_{(k,0)}$ and $\Sym_{(k,r)}$ should be
replaced by a Hilbert space, and such (massless) particles should be
described within principal series of infinite-dimensional
representations of the group $\spin_+(1,3)$.

$CPT$ groups of the tensor fields are constructed via the same
procedure that considered in the sections 4--6. For example, the
tensor field of the spin $1/2$ corresponding to the interlocking
scheme
\[
\dgARROWLENGTH=2.5em
\begin{diagram}
\node{\left(\frac{3}{2},1\right)}\arrow{e,<>}
\node{\left(1,\frac{3}{2}\right)}
\end{diagram}
\]
(which is equivalent to $(1/2,0)\oplus(0,1/2)$), is constructed
within the algebra
\begin{equation}\label{Bas12}
\C_2\otimes\C_2\otimes\C_2\bigotimes\overset{\ast}{\C}_2\bigoplus\overset{\ast}{\C}_2\otimes\overset{\ast}{\C}_2
\bigotimes\C_2\otimes\C_2\otimes\C_2.
\end{equation}
This algebra induces the spinspace
\[
\dS_2\otimes\dS_2\otimes\dS_2\bigotimes\dot{\dS}_2\otimes\dot{\dS}_2\bigoplus\dot{\dS}_2\otimes\dot{\dS}_2
\bigotimes\dS_2\otimes\dS_2\otimes\dS_2\simeq\dS_{64}.
\]
The spinbasis of the algebra (\ref{Bas12}) is defined by the
following $64\times 64$ matrices:
\[
\cE_1=\sigma_1\otimes\boldsymbol{1}_2\otimes\boldsymbol{1}_2\otimes\boldsymbol{1}_2\otimes\boldsymbol{1}_2
\otimes\boldsymbol{1}_2,\quad
\cE_2=\sigma_3\otimes\sigma_1\otimes\boldsymbol{1}_2\otimes\boldsymbol{1}_2\otimes\boldsymbol{1}_2\otimes\boldsymbol{1}_2,
\]
\[
\cE_3=\sigma_3\otimes\sigma_3\otimes\sigma_1\otimes\boldsymbol{1}_2\otimes\boldsymbol{1}_2\otimes\boldsymbol{1}_2,\quad
\cE_4=\sigma_3\otimes\sigma_3\otimes\sigma_3\otimes\sigma_1\otimes\boldsymbol{1}_2\otimes\boldsymbol{1}_2,
\]
\[
\cE_5=\sigma_3\otimes\sigma_3\otimes\sigma_3\otimes\sigma_3\otimes\sigma_1\otimes\boldsymbol{1}_2,\quad
\cE_6=\sigma_3\otimes\sigma_3\otimes\sigma_3\otimes\sigma_3\otimes\sigma_3\otimes\sigma_1,
\]
\[
\cE_7=\sigma_2\otimes\boldsymbol{1}_2\otimes\boldsymbol{1}_2\otimes\boldsymbol{1}_2\otimes\boldsymbol{1}_2
\otimes\boldsymbol{1}_2,\quad
\cE_8=\sigma_3\otimes\sigma_2\otimes\boldsymbol{1}_2\otimes\boldsymbol{1}_2\otimes\boldsymbol{1}_2\otimes\boldsymbol{1}_2,
\]
\[
\cE_9=\sigma_3\otimes\sigma_3\otimes\sigma_2\otimes\boldsymbol{1}_2\otimes\boldsymbol{1}_2\otimes\boldsymbol{1}_2,\quad
\cE_{10}=\sigma_3\otimes\sigma_3\otimes\sigma_3\otimes\sigma_2\otimes\boldsymbol{1}_2\otimes\boldsymbol{1}_2,
\]
\[
\cE_{11}=\sigma_3\otimes\sigma_3\otimes\sigma_3\otimes\sigma_3\otimes\sigma_2\otimes\boldsymbol{1}_2,\quad
\cE_{12}=\sigma_3\otimes\sigma_3\otimes\sigma_3\otimes\sigma_3\otimes\sigma_3\otimes\sigma_2.
\]
The extended automorphism group $\Ext(\C_{12})$ can be derived from
this spinbasis via the same calculations that presented in the
sections 4--6.

\section{Summary}
We have presented a group theoretical method for description of
discrete symmetries of the fields
$\boldsymbol{\psi}(\balpha)=\langle
x,\fg\,|\boldsymbol{\psi}\rangle$, where $x\in T_4$ and
$\fg\in\spin_+(1,3)$, in terms of involutive automorphisms of the
subgroup $\spin_+(1,3)\simeq\SU(2)\otimes\SU(2)$. We have shown that
an extended automorphism group $\Ext(\C_n)$, where $\C_n$ is a
complex Clifford algebra, lead to $CPT$ groups of the fields
$\boldsymbol{\psi}(\balpha)=\langle
x,\fg\,|\boldsymbol{\psi}\rangle$ of any spin defined on the
representation spaces (spinspaces) of $\spin_+(1,3)$. We considered
in detail $CPT$ groups for the fields of the type $(l,0)\oplus(0,l)$
(for example, $(1/2,0)\oplus(0,1/2)$, $(1,0)\oplus(0,1)$ and
$(3/2,0)\oplus(0,3/2)$). Also we discussed $CPT$ groups for the
fields of tensor type and their relations to particles of the same
spin with different masses. It would be interesting to consider
extended automorphism groups $\Ext(\cl_{p,q})$, where $\cl_{p,q}$ is
a real Clifford algebra, defined on the real representations of
$\spin_+(1,3)$. It would be interesting also to consider $CPT$
groups for the fields $\boldsymbol{\psi}(\balpha)=\langle
x,\fq\,|\boldsymbol{\psi}\rangle$ on the de Sitter group, where
$x\in T_5$ and $\fq\in\spin_+(1,4)\simeq\Sp(1,1)$, and for the
fields $\boldsymbol{\psi}(\balpha)=\langle
x,\fc\,|\boldsymbol{\psi}\rangle$ on the conformal group, where
$x\in T_6$ and $\fc\in\spin_+(2,6)\simeq\SU(2,2)$. Our next paper
will be devoted to these questions.

\end{document}